\documentclass[longauth]{aa}
\usepackage[varg]{txfonts}
\usepackage{adjustbox,lipsum}

\usepackage{graphicx}   
\newcommand\T{\rule{0pt}{2.6ex}}       
\newcommand\B{\rule[-1.2ex]{0pt}{0pt}} 

\begin{document}

\title{WEAVE-StePS. A stellar population survey using WEAVE at WHT                                                                                                                                                            }
\titlerunning{WEAVE-StePS. A stellar population survey}

\author{A.~Iovino\inst{1}\thanks{angela.iovino@inaf.it}
\and B.~M.~Poggianti\inst{2}
\and A.~Mercurio\inst{3, 4}
\and M.~Longhetti\inst{1}
\and M.~Bolzonella\inst{5}
\and G.~Busarello\inst{3}
\and M.~Gullieuszik\inst{2}
\and F.~La~Barbera\inst{3}
\and P.~Merluzzi\inst{3}
\and L.~Morelli\inst{6, 1}
\and C.~Tortora\inst{3} 
\and D.~Vergani\inst{5}
\and S.~Zibetti\inst{7} 
\and C.~P.~Haines\inst{6, 1} 
\and  L.~Costantin\inst{8}
\and  F.~R.~Ditrani\inst{1, 9}
\and  L.~Pozzetti\inst{5}
\and J.~Angthopo\inst{1}
\and M.~Balcells\inst{10, 11, 12}
\and S.~Bardelli\inst{5}
\and C.~R.~Benn\inst{10}
\and M.~Bianconi\inst{13}
\and L.~P.~Cassar\`a\inst{14} 
\and E.~M.~Corsini\inst{15,2}
\and O.~Cucciati\inst{5}
\and G.~Dalton\inst{16, 17}
\and A.~Ferr\'e-Mateu\inst{11, 12} 
\and M.~Fossati\inst{9, 1}
\and A.~Gallazzi\inst{7}
\and R.~Garc\'ia-Benito\inst{18}
\and B.~Granett\inst{1}
\and R.~M.~Gonz\'alez Delgado\inst{18}
\and A.~Ikhsanova\inst{15}
\and E.~Iodice\inst{3}
\and S.~Jin\inst{16,19,20} 
\and J.~H.~Knapen\inst{11, 12}
\and S.~McGee\inst{13}
\and A.~Moretti\inst{2}
\and D.~N.~A.~Murphy\inst{21}
\and L.~Peralta de Arriba\inst{21, 8} 
\and A.~Pizzella\inst{15, 2}
\and P.~S\'anchez-Bl\'azquez\inst{22, 23}
\and C.~Spiniello\inst{16, 3}
\and M.~Talia\inst{24, 5} 
\and S.~C.~Trager\inst{19}
\and A.~Vazdekis\inst{11, 12}
\and B.~Vulcani\inst{2}
\and E.~Zucca\inst{5}
}

\institute {INAF - Osservatorio Astronomico di Brera, via Brera 28, I-20121 Milano, Italy 
\and INAF – Osservatorio Astronomico di Padova, Vicolo dell’Osservatorio 5, I-35122 Padova, Italy 
\and  INAF - Osservatorio Astronomico di Capodimonte, Via Moiariello 16, I-80131 Napoli, Italy  
\and  Dipartimento di Fisica “E.R. Caianiello”, Università degli studi di Salerno, Via Giovanni Paolo II 132, I-84084 Fisciano (SA) 
\and  INAF – Osservatorio di Astrofisica e Scienza dello Spazio, Via P. Gobetti 93/3, I-40129 Bologna, Italy 
\and  Instituto de Astronomía y Ciencias Planetarias de Atacama (INCT), Universidad de Atacama, Copayapu 485, Copiap\'o, Chile  
\and INAF - Osservatorio Astrofisico di Arcetri, Largo Enrico Fermi 5, I-50125 Firenze, Italy  
\and Centro de Astrobiolog\'{\i}a (CSIC-INTA), Ctra de Ajalvir km 4, Torrej\'on de Ardoz, E-28850, Madrid, Spain 
\and Università degli studi di Milano-Bicocca, Piazza della scienza, I-20125 Milano, Italy 
\and Isaac Newton Group of Telescopes, ING, 38700 La Palma (S.C. Tenerife), Spain 
\and Instituto de Astrofísica de Canarias, IAC, Vía Láctea s/n, E-38205, La Laguna (S.C. Tenerife), Spain 
\and Departamento de Astrofísica, Universidad de La Laguna, E-38206, La Laguna (S.C. Tenerife), Spain 
\and School of Physics and Astronomy, University of Birmingham, Birmingham, B15 2TT, UK 
\and INAF - IASF Milano, via Bassini 15, I-20133 Milano, Italy 
\and Dipartimento di Fisica e Astronomia ``G. Galilei'', Universit\`a di Padova, vicolo dell'Osservatorio 3, I-35122, Padova, Italy 
\and Dept. Physics, University of Oxford, Keble Road, Oxford OX1 3RH, U.K. 
\and RAL, Space, Science and Technology Facilities Council, Harwell, Didcot OX11 0QX, U.K. 
\and Instituto de Astrof\'isica de Andaluc\'ia (CSIC), P.O. Box 3004, E-18080, Granada, Spain 
\and Kapteyn Astronomical Institute, Rijksuniversiteit Groningen, Landleven 12, 9747\, AD Groningen, the Netherlands 
\and SRON - Netherlands Institute for Space Research, Landleven 12, 9747\, AD Groningen, the Netherlands 
\and Institute of Astronomy, University of Cambridge, Madingley Road, Cambridge CB3 0HA, U.K. 
\and Departamento de F\'isica de la Tierra y Astrof\'isica, Universidad Complutense de Madrid, E-28040 Madrid, Spain 
\and Instituto de F\'isica de Part\'iculas y del Cosmos (IPARCOS), Universidad Complutense de Madrid, E-28040 Madrid, Spain 
\and Università di Bologna - Department of Physics and Astronomy, via Gobetti 93/2, I-40129, Bologna, Italy 
}

\date{Received November 3, 2022; accepted January 18, 2023}

\vspace{4cm}

\abstract 
{The upcoming new generation of optical spectrographs on four-meter-class telescopes will provide valuable
opportunities for forthcoming galaxy surveys through their huge multiplexing capabilities, excellent spectral resolution, and unprecedented wavelength coverage.}
{WEAVE is a new wide-field spectroscopic facility mounted on the 4.2 m William Herschel Telescope in La Palma. WEAVE-StePS is one of the five extragalactic surveys that will use WEAVE during its first five years of operations. It will observe galaxies using WEAVE MOS ($\sim$ 950 fibres distributed across a field of view of $\sim$ 3 square degrees on the sky) in low-resolution mode ($R\sim$ 5000, spanning the wavelength range $3660-9590$ \AA).}
{WEAVE-StePS will obtain high-quality spectra ($S/N \sim 10\,\AA^{-1}$ at $R \sim$ 5000) for a magnitude-limited ($I_{AB} = 20.5$) sample of $\sim 25,000$ galaxies, the majority selected at $ z \geq 0.3$. The survey goal is to provide precise spectral measurements in the crucial interval that bridges the gap between LEGA-C and SDSS data. 
The wide area coverage of $\sim$ 25 square degrees will enable us to observe galaxies in a variety of environments. The ancillary data available in each of the observed fields (including X-ray coverage, multi-narrow-band photometry and spectroscopic redshift information) will provide an environmental characterisation for each observed galaxy.}
{This paper presents the science case of WEAVE-StePS, the fields to be observed, the parent catalogues used to define the target sample, and the observing strategy that was chosen after a forecast of the expected performance of the instrument for our typical targets.}
{WEAVE-StePS will go back further in cosmic time than SDSS, extending its reach to encompass more than $\sim 6$ Gyr. This is nearly half of the age of the Universe. 
The spectral and redshift range covered by WEAVE-StePS will open a new observational window by continuously tracing the evolutionary path of galaxies in the largely unexplored intermediate-redshift range.
}

\keywords{galaxies: general - galaxies: formation - galaxies: evolution - galaxies: statistics
- galaxies: star formation - galaxies: stellar content}   

\maketitle




\section{Introduction \label{sec:introduction}}

One of the unfulfilled goals of astrophysics is to understand the physical processes that cause the formation and evolution of luminous structures to deviate so markedly from the assembly history of dark matter (DM) haloes. The large-scale structure evolution of the Universe is theoretically understood and well reproduced by simulations on scales larger than $\sim1$ Mpc. At these scales, the behaviour of the main agents that shape large-scale structure growth (gravity and dark energy) is relatively well known. At smaller scales, the stellar mass growth of galaxies within DM haloes, and thus the complex mechanisms that ae known as {\it gastrophysics} \citep{Bond1993}, depend on highly nonlinear processes (star formation, energetic feedback, and mergers, to mention a few) whose working details are still largely unknown \citep{Kuhlen2012, BullockBoylan-Kolchin2017}. 

Advanced theoretical models are needed to shed light on the mechanisms that regulate the connection between galaxies and their DM haloes and, in turn, on the connection with the surrounding large-scale structures. Still, the complexity of the problem is such that observations are essential to empirically constrain both theories and simulations of galaxy assembly history \citep{WechslerTinker2018}. 

In the past decades, the study of galaxies in the local Universe has greatly enriched our knowledge and understanding of galaxy evolution. 
Theoretical and empirical approaches are now anchored at $z\sim0$ by the large, uniform, and complete spectroscopic data of the Sloan Digital Sky Survey \citep[SDSS;][]{York2000}.  Observations of colours, morphology, spectral type, and star formation of galaxies show a striking bimodality in the galaxy population, where blue star-forming late-type galaxies are separated from red quiescent early-type galaxies \citep{Kauffmann2003, Blanton2003, Baldry2004}.  

The stellar mass of a galaxy is a powerful and reliable proxy for its total mass, and it is a fundamental quantity for predicting the position of each galaxy in the observed bimodal distribution. However, especially at lower galaxy stellar masses, the role of the environment in defining galaxy properties, and thus their position in the observed bimodal distribution, cannot be neglected \citep{Kauffmann2004, Baldry2006, BlantonMoustakas2009, Bamford2009}. 

The bimodality is also visible in spatially resolved observations of local galaxies \citep{Zibetti2017, LopezFernandez2018}, suggesting that its local and structural origin might lie within the galaxies. 
This bimodal distribution in galaxy properties persists at higher redshifts, where only the relative fraction of the two
populations in the bimodality peaks changes, while the role played by the environment becomes progressively less important \citep{Peng2010, Bolzonella2010, Iovino2010, Haines2017}. 

Parallel information on the global cosmic star formation history (SFH) provides an integrated view of galaxy evolution: there is firm evidence for a first phase of an increasing cosmic star formation rate (SFR) density in the Universe, reaching a peak at $z\sim2$, roughly 3.5 Gyr after the Big Bang, followed by a second phase in which the cosmic SFR density declines steadily to the present-day value \citep{MadauDickinson2014}.   Although largely independent of the complex evolutionary phases of each galaxy, this observational constraint provides a boundary that the cumulative galaxy SFHs need to satisfy. 

Many works have explored the fossil record information encoded in the spectra of local Universe galaxies, providing important constraints on the formation histories of galaxies of different masses, the {\it archaeological reconstruction}. However, retrieving information on the early phases of the SFHs from nearby galaxies has intrinsic limits: most local galaxies are too old for the differences in their early SFHs to be resolved because stellar spectra in the age range $> 5$ Gyr are very similar \citep[see][]{Gallazzi2005}. 

Galaxy spectra contain the luminosity-weighted output from billions of stars and the surrounding interstellar medium (ISM) and thus encode essential information on the physical properties of stellar populations and the gas content within each galaxy. In turn,  the most powerful way of probing galaxy evolution \citep[see][for a review]{Conroy2013} and of exploring the physical mechanisms that stop or rekindle star formation is tracing galaxy masses, stellar ages, chemical abundances, and the stellar initial mass function (IMF). 
 
An alternative approach consists of making a census of galaxies at increasingly earlier redshifts, which is called the {\it look-back approach}. By comparing snapshots of galaxy samples taken at different cosmic epochs, we can trace the evolution of galaxies as a population back in time. However, it is difficult to directly link the galaxies observed in the local Universe with their progenitors at higher redshifts because we are fundamentally ignorant of the actual progenitors of present-day galaxies of different types.  

The challenge for new observational studies is thus to combine the {\it archaeological} and the {\it look-back} approaches to provide precise measurements of a variety of physical parameters of galaxies, continuously and homogeneously sampled as a function of cosmic time and for a statistically robust sample of galaxies in different environments. The goal is to capture the richness (including the inherently stochastic nature) of the various processes involved in galaxy evolution, and to determine the different mechanisms (both internal and external) capable of modulating or shutting down star formation activity in galaxies and thus give rise to the observed galaxy bimodality in its many facets and its evolution. 

So far, the only notable survey designed to obtain high-quality high-resolution spectra in the relatively distant Universe is the Large Early Galaxy Census survey (LEGA-C), whose data are limited to the redshift window $0.6 \leq z \leq 1.0$ and to a sample of $\sim$ 3200 galaxies within the COSMOS field. The LEGA-C spectra have a median signal-to-noise ratio ($S/N$) of $\sim$20 $\AA^{-1}$ and a resolution $R \sim2500$.  They cover the observed wavelength range [6000 - 8800] $\AA$, corresponding to a rest-frame wavelength range [3750 - 5500]/[3000 - 4400] $\AA$ at $z \sim0.6/1.0$, respectively \citep[see][]{vanderWel2016, Straatman2018, vanderWel2021ApJS}. The LEGA-C observations have shown the full power of high-resolution continuum spectroscopy with a high signal-to-noise ratio as a tool for constraining galaxy formation models \citep{Bezanson2018, Wu2018, Wu2020, Chauke2018, Chauke2019}. 

However, fundamental questions remain open. The LEGA-C observations suggest that even the evolution of the most massive and supposedly passive galaxies does not always progress in a state of permanent quiescence after their initial star formation burst, as previously thought. Some of these galaxies may experience new star formation episodes between $z\sim0.8$ and $z\sim0$ \citep{Chauke2018, Wu2021}. Defining the mechanisms that give rise to this unexpected rekindling of star formation activity will provide important constraints on the evolutionary path of galaxies. 

\begin{figure*}[t]
        \centering
        \includegraphics[width=180mm]{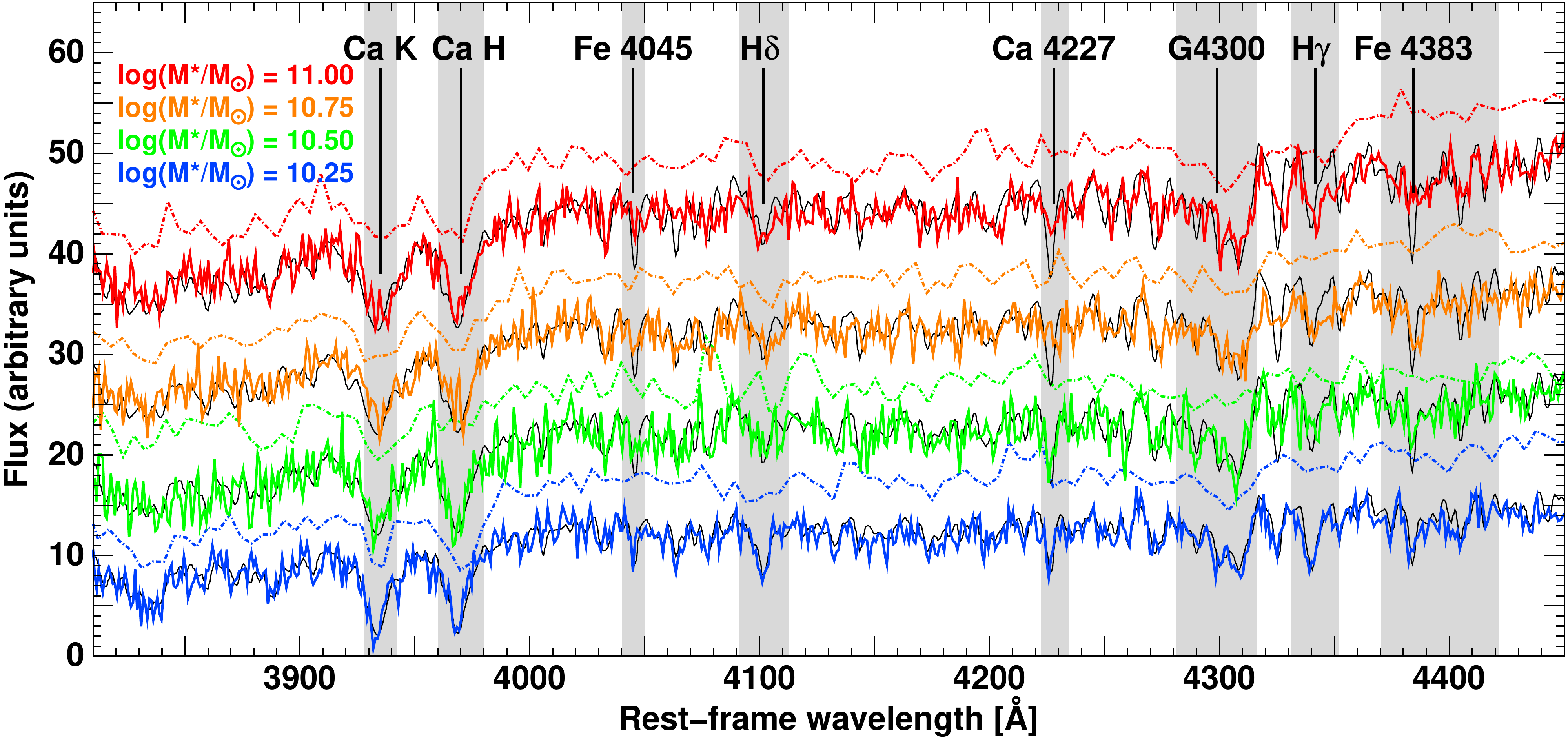}
        \caption{Comparison of typical SDSS ($z\sim0.1$, solid
                coloured lines) and VIPERS ($z\sim0.5$, dash-dotted lines) spectra
                of four early-type galaxies of different stellar masses as indicated in the plot. 
                VIPERS spectra are offset vertically by 10 units for
                clarity. The $S/N$ values of all these spectra are close to the median
                value for galaxies of that mass in the corresponding
                survey. The black curves show model spectra of a single stellar
                population with solar metallicities and ages in the range of 2.5-10 Gyr
                (from bottom to top), obtained using the MILES spectral library
                \citep{Vazdekis2010} and chosen to reproduce the data. Key spectral indices used for estimating
                the stellar ages or metallicities are indicated with their names, and the
                grey bands refer to the width of the Lick
                wavebands. The low $S/N$ and low resolution ($\sim$ 16 {\AA} FWHM, $R\sim 210$) of the
                VIPERS spectra smooth out (or fill in) the key spectral
                absorption features, preventing reliable
                measurements of the Lick indices for individual galaxies. The
                higher-resolution SDSS spectra retain the details of these absorption features.  
        }
        \label{fig:Figure1}
\end{figure*}

Surprisingly, the intermediate-$z$ window $0.3\leq z \leq 0.7$ remains largely unexplored by surveys of the quality required to adopt the archaeological look-back approach, limiting our ability to retrace the evolutionary path of galaxies without discontinuities over the $\sim 8$ Gyr since $z \sim 1$.
Until recently, only a few works have analysed ages and metallicities from high $S/N$ spectra of intermediate-redshift galaxies, most of them focusing on quiescent galaxies in clusters (e.g. \citet{Kelson2000, Sanchez-Blazquez2009, Jorgensen2013, FerreMateu2014, Werle2022}, but see \citet{Gallazzi2014} for an analysis of both quiescent and star-forming field galaxies). Despite the small sample sizes (a few tens to one hundred galaxies), these works have highlighted the importance of tracing the physical properties of galaxies at different redshifts and in different environments. 

The WHT Enhanced Area Velocity Explorer (WEAVE), a new optical spectrograph mounted at the $4.2$m William Herschel Telescope (WHT) in La Palma, offers the possibility of filling this gap in our knowledge.  Some of the unique capabilities of WEAVE are its wide field of view ($\sim 3$ square degrees on the sky), its high multiplexing capabilities ($\sim950$ fibres per pointing when used in MOS, i.e. multi-object spectrograph mode, with a minimum separation of fibres on the sky of $\sim60 \arcsec$), and its wide spectral coverage, $\sim3700$ to $\sim9600 \AA$ when used in its low-resolution mode at $R \sim$ 5000 \citep{Balcells2010, Dalton2012, Dalton2014, Dalton2016a}. The WEAVE LR mode provides an FWHM $\sim 1\AA$ at 5000 $\AA$, spanning values that vary from 0.9 $\AA$ in the blue arm and up to 1.5 $\AA$ in the red arm. This roughly corresponds to 3 pixels throughout the entIre covered wavelength range. 

The WEAVE-Stellar Population Survey (WEAVE-StePS in short) is one of the eight surveys (five of which are extragalactic) that will be performed by the WEAVE spectroscopic facility, using a total of approximately 1150 nights over five years of WHT time \citep[see][for more details on the planning of WEAVE operations]{Jin2022}. WEAVE-StePS will use $\sim3\%$ of this total budget of nights and aims to take full advantage of the WEAVE capabilities to obtain high-quality ($S/N \sim$ 10 $\AA^{-1}$ at $R \sim$ 5000) spectra of $\sim 25,000$ galaxies, the vast majority in the redshift range $0.3\leq z \leq 0.7$. The goal is to provide precise spectral measurements in the crucial redshift interval that bridges the gap between LEGA-C and SDSS data, thus rising to the challenge of characterising galaxy evolution in this intermediate-redshift range by obtaining data with a quality that is comparable to SDSS and LEGA-C data.  

In this paper, we describe WEAVE-StePS in detail. The paper is organised as follows. 
In Sect.~\ref{sec:GenGoals} we illustrate that the characteristics of the WEAVE spectrograph can provide a statistical galaxy sample of unique quality, enabling us to address significant questions about galaxy evolution. 
In Sect.~\ref{sec:SurveyPlan} we detail the choices that define WEAVE-StePS, the apparent magnitude and photometric redshift selection, and the specific fields that will be targeted by our survey.   
In Sect.~\ref{sec:WeaveObs} we discuss the WEAVE-StePS observational strategy, including the expected $S/N$ distribution for the galaxy sample observed by WEAVE-StePS. 
Finally, in Sect.~\ref{sec:Conclusions} we briefly summarize the main points that define WEAVE-StePS. 

We adopt $H_0=69.6$ km s$^{-1}$ Mpc$^{-1}$, $\Omega_M=0.286$,  and $\Omega_{\Lambda}=0.714$ as fiducial cosmological parameters throughout our paper \citep{Bennett2014}. All magnitudes are quoted in the AB system \citep{Oke1974}.  


\section{Main Science Goals of WEAVE-StePS \label{sec:GenGoals}} 

Three crucial ingredients are needed to shed light on the physical mechanisms that cause the different evolutionary routes of galaxies throughout cosmic time. 

\begin{figure*}
        \includegraphics[width=180mm]{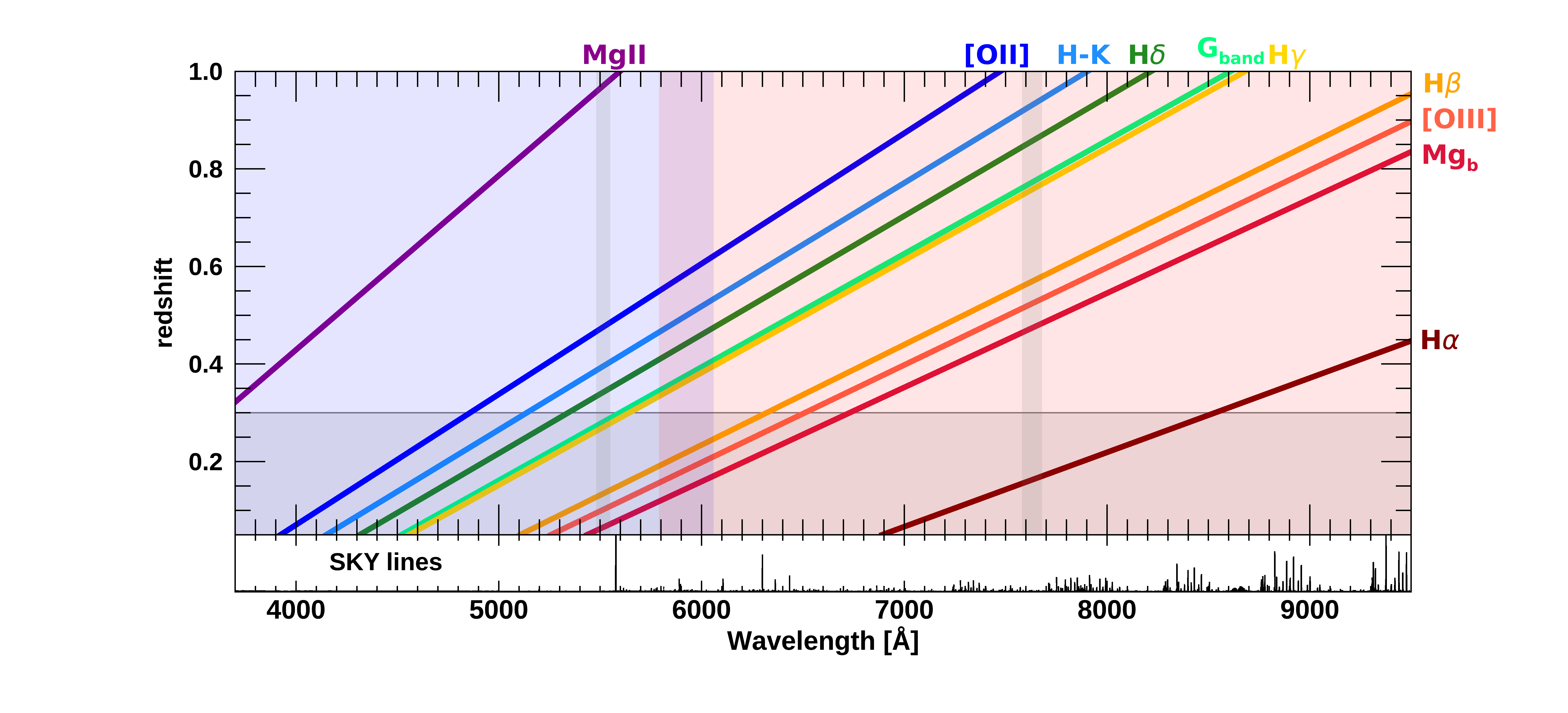}
        \caption{Observability of the main spectral indices as a function of redshift ($0.05<z<1.0$) in the
                wavelength range covered by the WEAVE spectrograph in its low-resolution mode.  
                The visibility of other emission/absorption lines is easily extrapolated from
                adjacent indices. The blue (red) region indicates the wavelength range covered by the blue (red) arm of the WEAVE spectrograph, and the overlap region covered by both arms is shaded in violet. The two narrower vertical grey bands are CCD gaps in the blue
                and red arm of the WEAVE spectrograph. They indicate the redshift ranges in which some of the features will not be available in observations.
                The empty region shows the redshift range in which the large majority of StePS targets will be
                observed ($z \geq 0.3$). The bottom panel shows an OH sky lines spectrum for reference.}
        \label{fig:Figure2}
\end{figure*}

The first ingredient is a large and well-defined sample of target galaxies, chosen to encompass a range of galaxy properties (stellar masses, morphologies, and colours, to mention the main properties) in a wide range of redshifts beyond the local Universe and located in a variety of environments. 

The second ingredient is high-quality photometric data for each target galaxy (the wider the wavelength range covered, the better). High-quality imaging enables the retrieval of  morphological parameters for each galaxy.  The photometric information in many bands enables the estimation of accurate photometric redshifts for each target galaxy (typical error in redshift ${\delta}_z \sim 0.03\times(1+z)$). With spectral energy distribution (SED) fitting techniques, rest-frame physical properties are retrieved: galaxy stellar masses, rest-frame magnitudes/colours, and, although with large uncertainties, first-order estimates of galaxy star formation rates, ages, and metallicities.

More accurate redshift information (e.g. those obtained from narrow-band filter photometry or from redshift surveys, with a typical velocity error ranging from ${\delta}_v \sim 100-1000$ km/s) is a valuable addition: it allows a more robust estimate of quantities from SED fitting. Furthermore, if the number density of galaxies with accurate redshift information is high enough in the considered volume, each galaxy can be assigned its precise location within the cosmic web, adding the environment variable to the analysis. 

The third and crucial ingredient is high-quality spectroscopic data for each target galaxy, meaning spectra with an adequate $S/N$ and spectral resolution over a sizeable rest-frame wavelength range, to obtain key physical information such as the age, metallicity, and abundance ratios of each galaxy stellar populations and ISM, and the kinematics for its stellar and gaseous components.  

Many large (N$_{\rm gal} \geq$ 10000) surveys of distant galaxies ($z \geq 0.3$) satisfy the first two requirements. To mention a few, theses are the AGN and Galaxy Evolution Survey \citep[AGES,][]{Kochanek2012}, the Smithsonian Hectospec Lensing Survey \citep[SHELS,][]{Geller2014}, the Galaxy and Mass Assembly survey \citep[GAMA,][]{Driver2012}, or the Baryon Oscillation Spectroscopic Survey \citep[BOSS,][]{Dawson2013} at the lower redshift end and the DEEP Extragalactic Evolutionary Probe 2 \citep[DEEP2,][]{Newman2013}, the VIMOS Public Extragalactic Redshift Survey \citep[VIPERS,][]{Guzzo2014}, zCOSMOS \citep{Lilly2009}, and the VIMOS-VLT Deep Survey \citep[VVDS,][]{Lefevre2015} at higher redshifts.
The spectral quality of these surveys (both their typical low $S/N$ values and low resolution, $R\sim500-1000$) makes them very efficient at measuring redshifts, but it is insufficient for retrieving galaxy stellar population and gas properties. 

Overcoming the limits of low $S/N$ spectroscopy by stacking spectra of a large number of galaxies selected in homogeneous bins (e.g. stellar mass, redshift, and velocity dispersion) is not a fully satisfying solution: it may provide the correct (light-weighted) average properties in a given bin, but misses the information about the scatter around this average and the higher moments of the distribution. Determining these moments is important for quantifying the number of stochastic events involved in galaxy evolution processes and the timescales of transitional phases. 

Only deep and high-resolution ($R\gtrsim 2000$) spectroscopy can provide reliable measurements of the high number of absorption features in the stellar continuum and the emission features superimposed on it, unlocking the variety of physical information that galaxy spectra can provide. Figure~\ref{fig:Figure1}, which directly compares VIPERS and SDSS spectral data, makes a clear case for this point. The low $S/N$ and low-resolution VIPERS spectra ($\sim$ 16 {\AA} FWHM) smooth out (or fill in) the key spectral absorption features, preventing reliable measurements of the Lick indices for individual galaxies. Even the power of emission-line diagnostics in these spectra is severely limited to the strongest-lined galaxies, which are just the tip of the iceberg of the star-forming population. 

Using SDSS spectra, \citet{Gallazzi2005} have argued that an $S/N \sim$ 20 $\AA^{-1}$ is the threshold for reliable age and metallicity estimates based on absorption features. More recent results have shown that stellar ages, metallicities, velocity dispersions, and dust extinction
can be accurately recovered without significant systematic offsets down to $S/N \sim$ 10 $\AA^{-1}$, albeit with significant uncertainties \citep{Choi2014, Magris2015, Citro2016}. 

The WEAVE spectrograph will provide adequate spectral resolution ($R \sim$ 5000 in its low-resolution, LR, mode and $S/N$ for relatively bright targets while offering another important advantage: its wide spectral range coverage, spanning the region $3660-9590 \AA$. The blue spectral range observed by WEAVE will map the UV rest-frame down to $\sim2900$ {\AA} for galaxies at $z \sim0.3$, extending to $\sim2200$ {\AA} for galaxies at $z \geq 0.7$.  This spectral window provides information on hot stellar components of galaxies, such as the young stars that trace the most recent tail of galaxy star formation histories. This is a novelty of the WEAVE observations, as LEGA-C data, covering the observed wavelength range [6000 - 8800] $\AA$, start to probe the rest-frame region below $\sim$ 3200 $\AA$  only for the highest-redshift galaxies. 

As shown in Fig.~\ref{fig:Figure2}, this wide wavelength range up to redshift $z \sim 1$ yields a large number of absorption and emission features that are useful tracers of physical properties of the stellar and ionised-gas components. 
WEAVE spectra can provide homogeneous measurements of gas metallicity, and they can distinguish between AGN, LINER, and star-formation-powered emission using the same set of lines over a wide redshift range. The [OIII]$\lambda$5007 emission line and the H$_{\beta}$ line are both within the wavelength range of WEAVE up to z $\sim 0.9$, while $\rm H\alpha$ is available up to $z\sim0.4$ and can be used for different line-ratio diagnostics detection of AGN \citep[][and references therein]{Kewley2006, Lamareille2010, Juneau2011}.
Similarly, the [OII]$\lambda$3727 line is available as a star formation indicator and can be calibrated using $\rm H\alpha$ up to $z\sim0.4$. 
Metallicity indicators for refining the SFR estimate are available up to $z=0.7$, for example [OIII]5007/$H_{\beta}$, [OIII]5007/[OII]3727 and, when available, [NeIII]3870/[OII]3727 \citep{Nagao2006, KewleyEllison2008, Speagle2014}. 

\begin{figure*}
        \includegraphics[width=180mm]{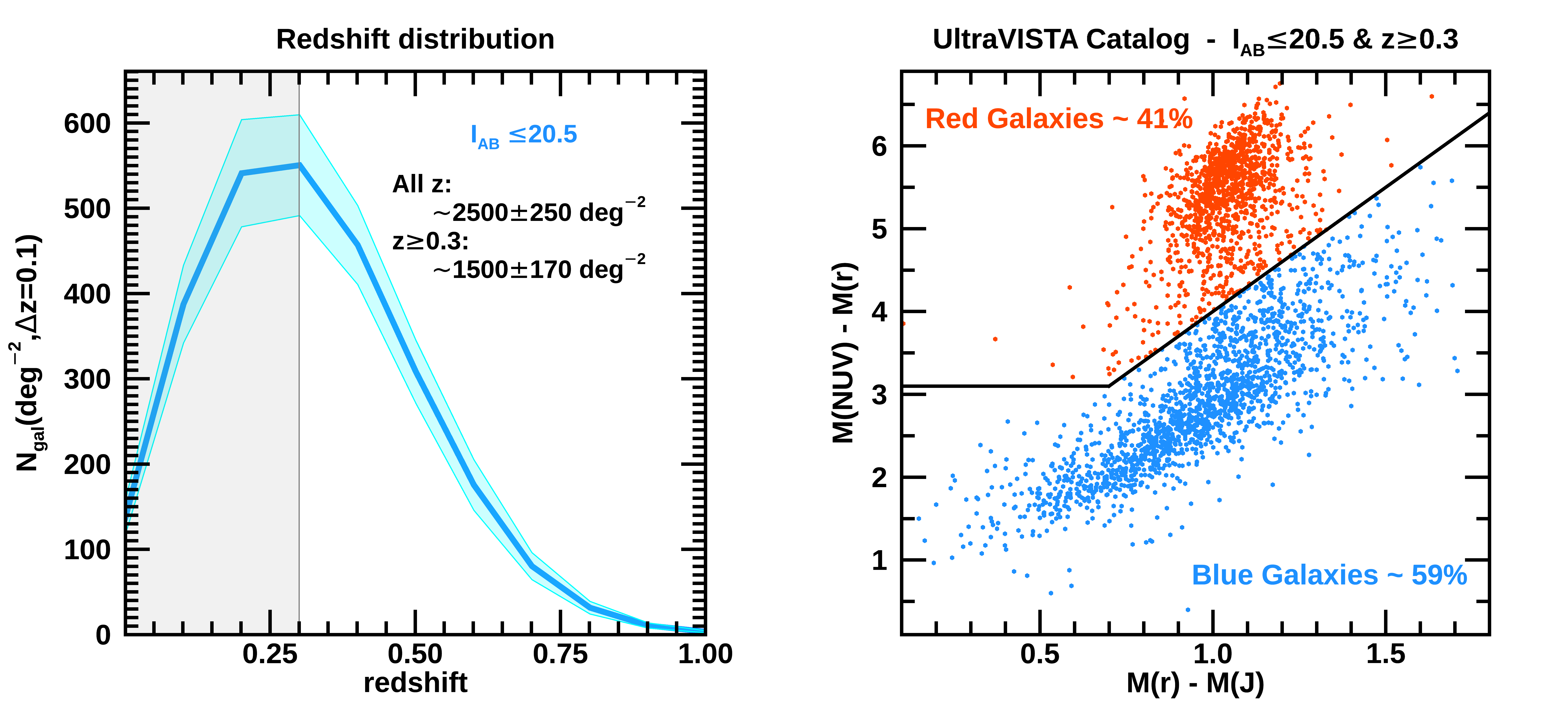} 
        \caption{Expected redshift and colors distribution of WEAVE-StePS galaxies. Left panel: Redshift distribution of a magnitude-limited 
                sample of galaxies down to $I_{AB} \sim20.5$. The y-axis
                shows the expected surface density of galaxies per square
                degree and per 0.1 redshift interval. The mean values (thick
                line) and their field-to-field variance (shaded region) were
                obtained in random positions within the CFHTLS-W1 and CFHTLS-W4 photometric catalogues
                \citep{Ilbert2006, Hudelot2012}. These 
                numbers were obtained by removing all regions
                masked because of lower-quality photometry.  
            Right panel: Expected two-colour distribution of our target sample on the rest-frame colour-colour $(NUV-r)$ and $(r-J)$ plane, as obtained using the WEAVE-StePS sample definition and the UltraVISTA catalogue in the COSMOS field \citep{McCracken2012, Ilbert2013}.
                This colour-colour plot is an efficient way to distinguish between quiescent and star-forming galaxies, avoiding a mix between 
            dusty star-forming galaxies and quiescent galaxies, as extinction moves star-forming galaxies along a diagonal direction parallel to the slanted line in the plot \citep[see ][for more details]{Ilbert2010}.
                } 
        \label{fig:Figure3}
\end{figure*}

The potential of WEAVE-like spectra for galaxy evolution studies has been investigated in detail in \citet{Costantin2019} and in Ditrani et al. (2023, {\it in prep}), using simulations that mimic WEAVE-expected performances for the full variety of spectral galaxy types that WEAVE-StePS will observe. 
Adopting Bayesian analysis, as developed in \citet{Gallazzi2014}, coupled with the extended galaxy model library of \citet{Zibetti2017}, \citet{Costantin2019} have shown that with the full suite of indices available to WEAVE, even with an $S/N$ as low as $\sim$ 10 ${\AA}^{-1}$ in the observed I band, it is possible to be sensitive to secondary episodes of star formation younger than $\sim 0.2$ Gyr in addition to a bulk stellar population older than $\sim$ 5 Gyr, which is the expected age for the stellar population of massive galaxies that WEAVE-StePS will target. It will thus be possible to recover evidence of past rejuvenation episodes in old galaxies, or conversely, to identify galaxies containing truly long-term passive old stellar populations.
In a parallel paper and using the same method, Ditrani et al. (2023, {\it in prep}) have shown that unbiased stellar metallicity estimates can be obtained in WEAVE-like spectra at an $S/N$ as low as $\sim$ 10 ${\AA}^{-1}$ in the observed I band. For these typical $S/N$ values, the uncertainties in the retrieved metallicity are high when the spectra are dominated by young stars ($\sim0.3$ dex for galaxies with $r$-band-weighted ages younger than 1 Gyr), but they decrease significantly for the older galaxy population (down to $\sim 0.2$ dex for galaxies with $r$-band-weighted ages older than 6 Gyr). Before we describe our survey planning in detail, we concisely list the principal physical information we intend to obtain for WEAVE-StePS galaxies:
\begin{itemize}
    \item {The age(s) of the stellar component}. 
WEAVE-StePS high-quality data will provide reliable estimates of stellar population ages. The distribution of stellar population ages will
provide constraints on the stellar mass growth of galaxies across cosmic time from $z\sim 1$ to today, a redshift range that will be available by combining the information provided by WEAVE-StePS with LEGA-C and SDSS data. 
\vspace{0.5mm}
\item {The star formation activity timescale}.  
The definition of the typical timescale for the decrease in star formation activity in galaxies and the parallel quenching phenomena will benefit from a detailed estimate of the stellar population ages of galaxies.  WEAVE-StePS spectra will also enable us to estimate the presence of rejuvenating events and, conversely, will provide the demographic of truly passive old galaxies \citep[see][]{Costantin2019}. 
\vspace{0.5mm}
\item {The metal abundances in stars and gas}.
We will use WEAVE-StePS spectra to measure individual chemical abundances in galaxy spectra \citep[see e.g. ][]{Labarbera2017, Gallazzi2021} and investigate the mass–metallicity relations for the gaseous and stellar components of observed galaxies, as well as the fundamental metallicity relation \citep{Mannucci2010}. 
This is an interesting indicator of galaxy evolution. It reflects gas cycling through stars and any gas exchange (inflows or outflows) between the galaxy and its environment \citep[see][for a comprehensive review]{Maiolino2019}. 
\vspace{0.5mm}
\item {The presence/absence of AGN activity}. 
The wide wavelength range covered by WEAVE-StePS spectra will allow us to use a common set of lines as AGN/LINER activity indicators to distinguish between AGN, LINER, and star-formation-powered emission. The AGN fraction demographic will shed light on the role of AGN in galaxy evolution \citep{Heckman2014}. 
\vspace{0.5mm}
\item {The  presence of inflows and outflows}. 
The phenomena of inflows and outflows of gas play an important role in regulating the growth of galaxies as powerful agents to accrete/remove gas from the galaxy \citep{Rubin2012, Harrison2018}. The [OIII]5007 emission line, which is used to detect and study AGN outflows \citep{Woo2016}, is within the wavelength range of WEAVE up to z $\sim$ 0.9. 
The adopted R $\sim$ 5000 spectral resolution, corresponding to a FWHM $\sim 50$ km/s, will enable WEAVE-StePS to detect the different kinematic components expected in these objects. This in turn will enable the separation of narrow ($\sim$ 100 km/s) and broad (>300 km/s) components that are generally present in AGN outflows.  
\vspace{0.5mm}
\item {The stellar and dynamical mass of galaxies}. 
An accurate estimate of stellar and dynamical masses is crucial to link the evolution of galaxies with their DM halo \citep{Tortora2018}. WEAVE spectra will enable estimates of galaxy velocity dispersion with a statistical accuracy below $\sim 15\%$ and with deviations smaller than $\sim30\%$ \citep{LaBarbera2010}. WEAVE-StePS will be the first survey to provide a large sample of galaxies in the intermediate-redshift range with accurate estimates of stellar velocity dispersions and dynamical and stellar masses. We will also be able to constrain the stellar IMF using gravity-sensitive features \citep[see e.g. ][]{LaBarbera2013, Spiniello2015} and its impact on stellar mass estimates at intermediate redshifts.
\vspace{0.5mm} 
\item {Green valley, or transition, galaxies}. 
We expect to observe a sizeable sample of transition/green-valley galaxies (see Figure \ref{fig:Figure3}). These galaxies are located in the wide region separating blue and red galaxies. They have abandoned intense star formation activity and are on their way to reaching a passive status. They contribute to the bending and broadening of the star-forming main sequence observed at the high galaxy mass ranges we target \citep{Popesso2019} and can help to shed light on the mechanisms that quench star formation in galaxies \citep{Angthopo2021}.  
\end{itemize} 

The large sample of galaxies that WEAVE-StePS plans to observe (roughly 25000; see Section \ref{sec:WeaveObs}) will provide an opportunity for detecting and selecting rare/peculiar targets. 
An interesting set of objects that may serendipitously enter our observations is that of strongly lensed galaxies, which are detected through their emission lines that are overlaid on the observed foreground galaxy spectra; see \citet{Warren1996} for a prototypical such case, and more recently, \citet{Bolton2006} and \citet{Talbot2021}. Our target sample will include many massive galaxies at $ \sim 0.5$, which are ideal lenses considering their mass and location in redshift. We can therefore expect a few lensed galaxies to enter our sample whenever foreground (lens) and background (lensed) galaxies are well aligned. 

Last but not least, WEAVE-StePS will undoubtedly be a legacy survey. Regular public data releases are planned for WEAVE data, with an expected yearly cadence of public data releases, following the first public data release two years after the start of survey operations \citep{Jin2022}. The spectra of thousands of WEAVE-StePS galaxies and their associated measurements (e.g.  emission lines and the equivalent widths of stellar absorption features) will be valuable for the identification and characterisation of galaxy populations in large sky surveys. As an example, the spectra and equivalent widths of emission lines such as $\rm H\alpha$, $H_{\beta}$, [NII]6585,  [OIII]5007, and [OII]3727 can be input to machine-learning algorithms applied to narrow-band photometric surveys such as J-PAS \citep{Benitez2014}.  This information will make it possible to infer the emission-line properties of objects through their observed narrow-band colours, enabling unbiased samples of QSO, AGN and star-forming galaxies to be identified up to z$ \sim$1 \citep{Martinez-Solaeche2021, Martinez-Solaeche2022}.  


\section{Survey plan of WEAVE-StePS \label{sec:SurveyPlan}} 

This section details the constraints that define the WEAVE-StePS galaxy sample and explore its redshift, colour, and stellar mass distributions. We then present the WEAVE-StePS target fields: their position, area coverage, and available ancillary data (both photometric and spectroscopic), and we discuss the quality of the astrometry for our input catalogues and that of the available photometric redshifts.  Finally, we conclude the section by discussing the yearly visibility of each of the chosen fields. 

\subsection{Sample selection: Magnitude and redshift constraints\label{subsec:selection}} 
One of the main characteristics of the WEAVE spectrograph is its high multiplexing capability. 
The sample of galaxies to be observed is expected to have a surface density that fully exploits this characteristic while covering a magnitude range that enables obtaining high $S/N$ spectra at $R\sim$ 5000 with a 4m class telescope.   

The two simple constraints that define the WEAVE-StePS target galaxy sample are:
\begin{itemize}
        \item target total observed magnitude $I_{AB} \leq 20.5$, and 
        \item target redshift $z \geq 0.3$.
\end{itemize}

As we show in Section \ref{sec:WeaveObs}, this magnitude limit is well within the capabilities of WEAVE to produce spectra at $S/N \sim 10\,\AA^{-1}$ for the majority of the targeted sample. 
In the following, we explore the distribution in redshift and the physical properties of the sample of galaxies defined using these two simple constraints, taking advantage of public catalogues available in three of the four fields that we will target (see Section \ref{subsec:FieldsChoice}), namely the two CFHTLS fields CFHTLS-W1 and CFHTLS-W4, and the COSMOS field. 

To assess the number density of the sample of galaxies defined by the two WEAVE-StePS constraints, we used the publicly available photometric catalogues (including photometric redshifts) in the two wide areas of the CFHTLS-W1 and CFHTLS-W4 fields \citep{Ilbert2006, Hudelot2012}, excluding galaxies that are located in poor-quality/masked regions. 
The resulting redshift distribution of target galaxies, including its field-to-field variance (i.e. the $rms$ of the counts obtained at different random sky positions within the two fields) is shown in the left panel of Fig.~\ref{fig:Figure3}. 

Roughly $\sim$60\% of the magnitude-selected targets are located at $z \ge$0.3, and only $\sim2.5\%$ are located at $z \ge$0.7. The actual cumulative numbers are  $\sim2500 \pm 250$ galaxies per square degree when no redshift pre-selection is applied, and $\sim1500\pm 170$ galaxies per square degree when the subset of galaxies at $z \geq$ 0.3 is considered. 

The constraint in redshift ($z \geq$ 0.3, photometric or spectroscopic whenever available) combined with the bright magnitude selection focuses the interest of WEAVE-StePS on the intermediate-redshift range between LEGA-C and SDSS: $0.3 < z <0.7$, with only a tail of the sample, roughly $4\%$, located at $z \ge$0.7. Only in the COSMOS field will we relax the redshift limit to $z \geq$ 0.1 the (see Section \ref{subsec:FieldsChoice} for more details). 

The two constraints in $I_{AB}$ magnitude and redshift imply that a total of $\sim$ 5000 galaxies are available for targeting within each WEAVE FoV pointing ($\sim$ 8000 if no redshift cut-off is chosen). This density of targets is very well suited to the multiplexing capabilities of WEAVE, enabling multiple passes on the same sky position (see Section \ref{sec:WeaveObs}). It is worth noting that these numbers represent a conservative lower limit for target surface densities that is obtained by removing lower-quality masked regions in any of the bands that are used to compute photometric redshifts, regions surrounding stars, and general problematic areas.  

To explore the physical properties of the galaxies selected by WEAVE-StePS in even more detail, we take advantage of the COSMOS field. This field covers a smaller area, but with a much larger set of ancillary data. 
The publicly available UltraVISTA catalogue provides high-quality photometric redshifts, rest-frame magnitudes, and galaxy stellar masses that were obtained using the code \texttt {LePhare} and a library of synthetic spectra generated using the Stellar Population Synthesis (SPS) model of Bruzual and Charlot \citep{BruzualCharlot2003} and a Chabrier IMF \citep{Chabrier2003} for each galaxy in the COSMOS field down to $\sim$ 24 in the $K_{sAB}$ band. 
The right panel of Fig.~\ref{fig:Figure3} shows  the rest-frame  $(NUV-r)$
vs $(r-J)$ colour-colour diagram for galaxies selected using WEAVE-StePS criteria in the UltraVISTA catalogue. This diagram is commonly used to distinguish between red/passive and blue/active galaxies \citep{Ilbert2013}.  It clearly illustrates that the galaxies targeted by WEAVE-StePS have an approximately equal percentage of red/blue galaxies ($41\%$ ($59\%$ respectively). The WEAVE-StePS selection criteria will sample the broad variety of galaxy types within the galaxy population, from blue star-forming to transition galaxies, and finally, to red and passive ones. 

\begin{figure}
        \includegraphics[width=90mm]{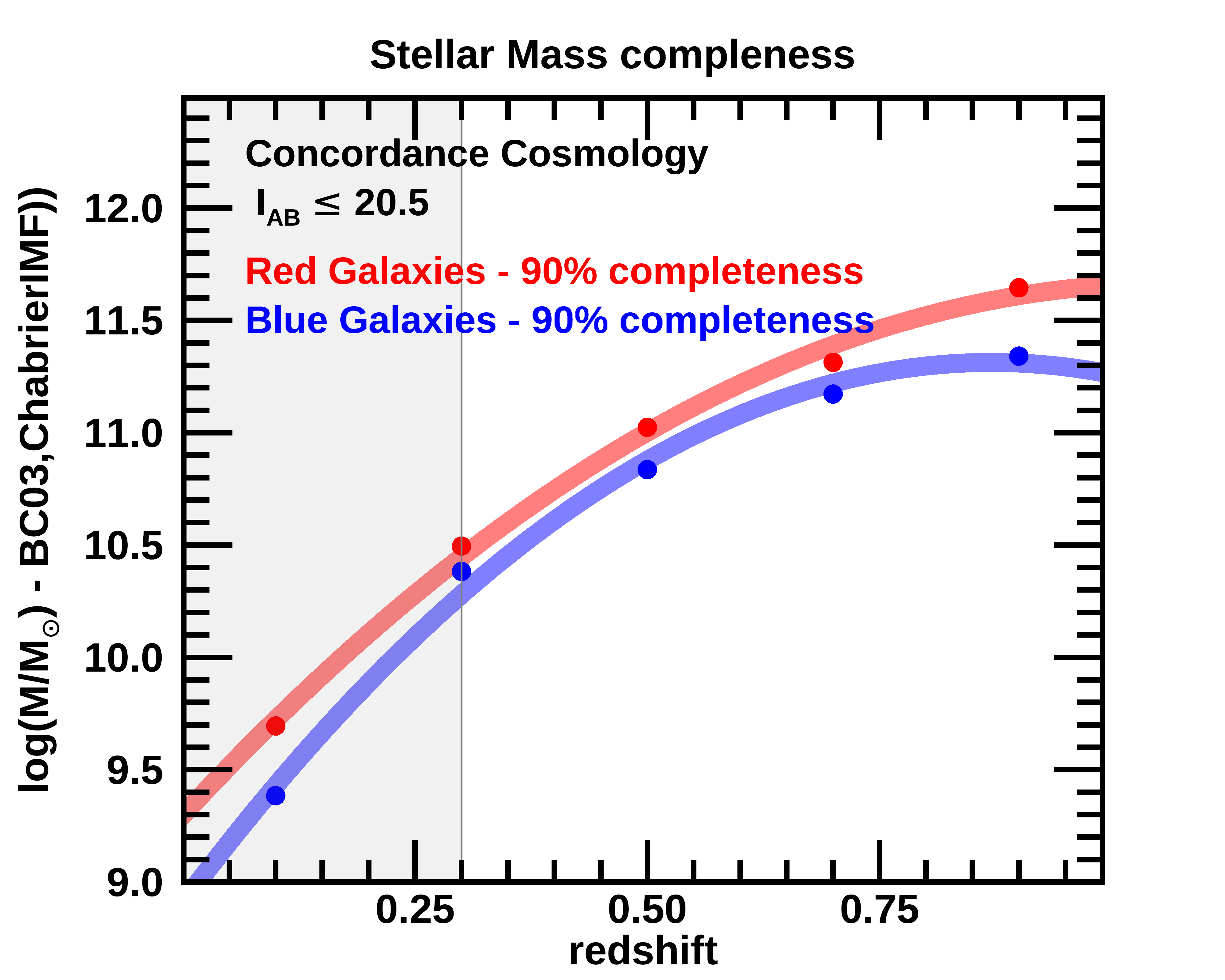} 
        \caption{ 90$\%$
                colour-dependent completeness limit in stellar mass for blue and red galaxies as
                a function of redshift (blue and red lines). We used data and stellar masses
                (Chabrier IMF) from the UltraVISTA catalogue in the COSMOS field
                \citep{McCracken2012, Ilbert2013}, and the blue and red galaxies were defined using a simple colour-colour rest-frame $NUV-R-J$ plot. 
                The shaded area indicates the $z\leq 0.3$ redshift range.} 
        \label{fig:Figure4}
\end{figure}

For both the red and the blue galaxy populations, Fig.~\ref{fig:Figure4} shows the estimated colour-dependent completeness limit in stellar mass. To obtain this, we used the galaxy stellar masses estimates provided in the UltraVISTA catalogue and a procedure similar to that described by \citet{Pozzetti2010}. We first calculated the stellar mass  that each target galaxy would have at its redshift if its apparent magnitude were the limiting magnitude of the survey $I_{AB} = 20.5$, and then we used these limiting stellar masses to define the stellar mass value below which $90\%$ of galaxies lie. We calculated this separately for the blue and red samples, obtaining the completeness values shown in  Fig.~\ref{fig:Figure4}, that is, an average typical stellar mass limit equal to log(${\cal M/M}_\sun$)$\sim 10.4/11.0/11.3$ at z $\sim$ 0.3/0.5/0.7. Thus we sample the high stellar mass tail of the galaxy population at all explored redshifts.  

These galaxies contribute significantly to the fraction of baryonic mass locked in stars, and their evolution is still far from complete at the redshifts targeted by WEAVE-StePS. 
The idea of massive galaxies as systems that formed all their stars early in time and are passively evolving since then has been challenged in the past few years: recent results have shown that the ages of massive galaxies (${\cal M} > 10^{11} {\cal M}_\sun$) at $z \sim0.8$ are inconsistent with
those of their local counterparts \citep{Wu2018, Spilker2018, Chauke2018} when a passive evolution of the whole massive galaxy population is assumed. This result is also indicated by the apparent slowing down of the $D4000$ value evolution of massive passive galaxies after $z\sim0.5$ (Haines et al. 2023, \emph{in prep.}). Our sample of massive galaxies at intermediate redshifts will provide observational constraints to shed light on this topic.  

\subsection{Sample selection: Choice of target fields\label{subsec:FieldsChoice}} 

The WEAVE spectrograph is fibre-fed. Each MOS science fibre has an 85 $\mu$m core, which subtends $1\farcs3$ on the sky, and an expected positioning accuracy $\sim0\farcs2$ rms \citep{Hughes2020}.  The target coordinates provided to the spectrograph must be accurate and precise down to similar levels. They should be defined on the Gaia reference frame of EDR3 \citep{Gaia2021}, the same as was adopted by the general WEAVE survey planning.
Thus the availability of quality astrometry is a crucial prerequisite when defining the target fields for WEAVE-StePS. 

As discussed in Sect.~\ref{sec:GenGoals}, the choice of target fields should be further driven by the availability of high-quality ancillary data, both photometric (the wider the wavelength range covered, the better) and spectroscopic. 
The presence of high-quality ancillary photometric data over a wide wavelength range will improve the quality of the photometric redshift information available for each target, an important parameter to consider as we have adopted a redshift cut-off in our survey design (see Sect.~\ref{sec:SurveyPlan}). 
The availability of a sizeable percentage of targets with spectroscopic redshift information is a further important bonus, as it enables us both to test and improve the quality of photometric redshifts and to refine environment characterisation for the targeted galaxies. 
Last but not least, the fields to be observed should be distributed evenly in $RA$ to permit observations with WEAVE throughout the year and be located in $Dec$ positions such that each field is positioned at relatively low air masses for a sizeable number of hours in the nights when it is available for observations. 

Our selection takes advantage of the recently released Hyper-Supreme-Cam Subaru Strategic Program (HSC-SSP) data \citep{Aihara2018a}. 
These data cover several fields that are all well accessible from the WHT, as the Subaru and William Herschel telescopes are at latitudes that differ by only $\sim$ 10 degrees. 
Each HSC-SSP field is observed in five broadband filters modelled on the SDSS filter set ($g,r,i,z,y$) at various depths (wide, deep, and ultradeep, depending on the sky position) and offers the important advantage of using the PanSTARRS1 DR2 catalogue as primary calibration source \citep{Schlafly2012, Magnier2013, Chambers2016}, a catalogue astrometrically calibrated against Gaia DR1 \citep{GAIA2016a, GAIA2016b}. The wide fields of HSC-SSP reach a $5\sigma$ I-band magnitude depth of $~26.2$ for point sources and a corresponding $5\sigma$ I-band limiting magnitude within apertures of two arcseconds in diameter that are shallower by $\sim 0.3$ mag. This depth is well beyond our selection needs. 

We used the available HSC-SSP PDR2 data \citep{Aihara2019}, incremental version $2$, as available when we produced our target catalogues \footnote{See \url{https://hsc-release.mtk.nao.ac.jp/doc/index.php/sample-page/pdr2/} for a full description}. Any subsequent changes in later releases \citep[see also][for a detailed list of new observations and improvements introduced in the third data release]{Aihara2022} were considered negligible for our purposes. From here onwards, when we quote HSC-SSP data, we mean DR2 data of incremental version $2$ . These data include target catalogues in the different bands, a list of masked regions, and photometric redshifts for each target.  

\begin{table}
        \caption{List of fields covered by WEAVE-StePS}
        \centering{
        \begin{tabular}{cccccc}
       \hline \hline 
                Name                     &        RA(J2000)     &         Dec(J2000)     \T\B  \\  
       \hline
                COSMOS                   &        10 00 30      &         +02 12 21       \\
            ELAIS-N1             &        16 11 00      &         +54 59 17       \\
                CFHTLS-W4            &    22 15 00      &         +01 36 00        \\
                CFHTLS-W1       &    02 24 00      &      $-$05 06 00        \\
       \hline
        \end{tabular}
        }
        \label{tab:fields}
\end{table}

In the following, we detail the available ancillary information for each of the four fields that WEAVE-StePS will target. The information is listed in Table \ref{tab:fields}. All magnitudes in our catalogues are corrected for galactic extinction and were computed using the newer estimates of Galactic dust extinction from \citet{Schlafly2012} and the mean extinction curve of the diffuse interstellar medium with $R_V$=3.1 from \citet{Fitzpatrick1999}. 

\subsubsection{COSMOS field\label{subsubsec:COSMOS}}  The COSMOS field is the first obvious choice as a WEAVE-StePS target field (see Table \ref{tab:fields} for its coordinates). COSMOS is still the largest of the classic extragalactic fields imaged by the HST: a $\sim1.7$ square-degree  mosaic with ACS in the F814W band \citep{Scoville2007}. This enables excellent structural parameter estimates and morphological classifications \citep{Sargent2007, Scarlata2007, Kartaltepe2015}. Forthcoming JWST observations are planned for the area. This field is covered by HSC-SSP data, with four extended-COSMOS pointings to deep depth, and a fifth pointing to ultradeep depth \citep{Aihara2018b}. These HSC data further enrich the ample dataset available for this field, which outside the optical data range includes two UV bands from the Galaxy Evolution Explorer (GALEX; \citet{Morrissey2005, Bianchi2011, Bianchi2017}), deep u-band imaging from the Canada–France–Hawaii Telescope \citep[CLAUDS; see][]{Sawicki2019}, YJHK broadband data from the VIRCAM instrument on the VISTA telescope, and the four IRAC/Spitzer channels \citep[see][for a complete list and depths of the available photometric data]{Laigle2016, Weaver2022}.  
This field has also been the target of various spectroscopic campaigns, most notably zCOSMOS \citep{Lilly2007}, the PRIsm MUlti-Object Survey \citep[PRIMUS][]{Coil2011}, and recently, the LEGA-C survey \citep{vanderWel2016}. In particular, the LEGA-C survey spectra will provide an important  control sample for the galaxies in common with WEAVE-StePS. Part of the area covered by WEAVE-StePS in COSMOS is covered by deep XMM and Chandra pointings \citep{Hasinger2007, Cappelluti2009, Elvis2009}. 
The COSMOS field is observable in the first trimester of the year and is the only one for which we decided to extend our target selection to lower redshifts (down to $z \gtrsim 0.1$) to explore the trends that the main WEAVE-StePS sample will observe to higher redshifts down to lower masses. 

\subsubsection{ELAIS-N1 field\label{subsubsec:ELAIS-N1}}  
The second WEAVE-StePS field is ELAIS-N1 (see Table \ref{tab:fields} for its centre), another well-known extragalactic field. This field was originally chosen for deep extragalactic observations with the Infrared Space Observatory (ISO) because the infrared background is low \citep{Oliver2000}. Since then, a variety of other photometric data has been added. 
The ELAIS-N1 field is covered by HSC-SSP four deep pointings, totalling $\sim 6.5$ square degrees. Ancillary data include deep CFHT Telescope U band data \citep[CLAUDS; see][]{Sawicki2019}, JK Near InfraRed (NIR) broadband UKIDSS-DSX data \citep{Lawrence2007, Swinbank2013}, Spitzer IRAC+MIPS data, and Herschel PACS+SPIRE data \citep[see][for a summary of the available bands and their depths]{Malek2018, Shirley2021}. It is worth noting that ELAIS-N1 is the deepest LOFAR two-metre sky survey field \citep{Sabater2021}. The number of spectroscopic redshifts in this field is somewhat limited and mostly consists of data from SDSS-DR16 \citep{Ahumada2020}, which has been complemented with the compilation of \citet{Vaccari2022}. ELAIS-N1 can be observed in the second trimester of the year.   

\subsubsection{CFHTLS-W4 and CFHTLS-W1 fields\label{subsubsec:CFHTLS}}  
The third and fourth fields are two areas within the CFHTLS-W4 and CFHTLS-W1 wide fields \citep{Hudelot2012}, both covered by HSC-SSP Wide data, and centred on the sky regions covered by spectroscopic VLT-VIMOS data.    
The ancillary photometric data for both CFHTLS-W4 and CFHTLS-W1 include the u-band photometry from the CFHTLS T07 data release \citep{Hudelot2012}, YJHK NIR broadband UKIDSS-LAS data \citep{Lawrence2007, Lawrence2012}, and the VIPERS Multi-Lambda Survey, which for part of the fields provides ready-to-use matched photometry from the GALEX Deep Imaging Survey, and NIR photometry in K$_{s}$ band obtained at WIRCam \citep{Moutard2016}. 
Galaxies in both fields possess morphological parameter estimates, both from the CFHTLenS collaboration \citep{Miller2013} and from the VIPERS collaboration \citep{Krywult2017}. The availability of morphological parameters, even though it is of lower quality than those provided by HST data in the COSMOS field, will be crucial to obtain reliable dynamical masses.

The CFHTLS-W4 field can be observed in the third trimester of the year, and the centre chosen for WEAVE-StePS is listed in Table \ref{tab:fields}.   The centre choice is such that this field largely overlaps the VIPERS-W4 spectroscopic data and therefore includes VIPERS observations at $z\geq0.5$ \citep{Guzzo2014, Scodeggio2018}. There is partial overlap with spectroscopic data from the VVDS-Wide survey \citep{Garilli2008} and from Stripe 82 of the SDSS Legacy Survey \citep{Annis2014}.    

\begin{figure*}
        \includegraphics[width=180mm]{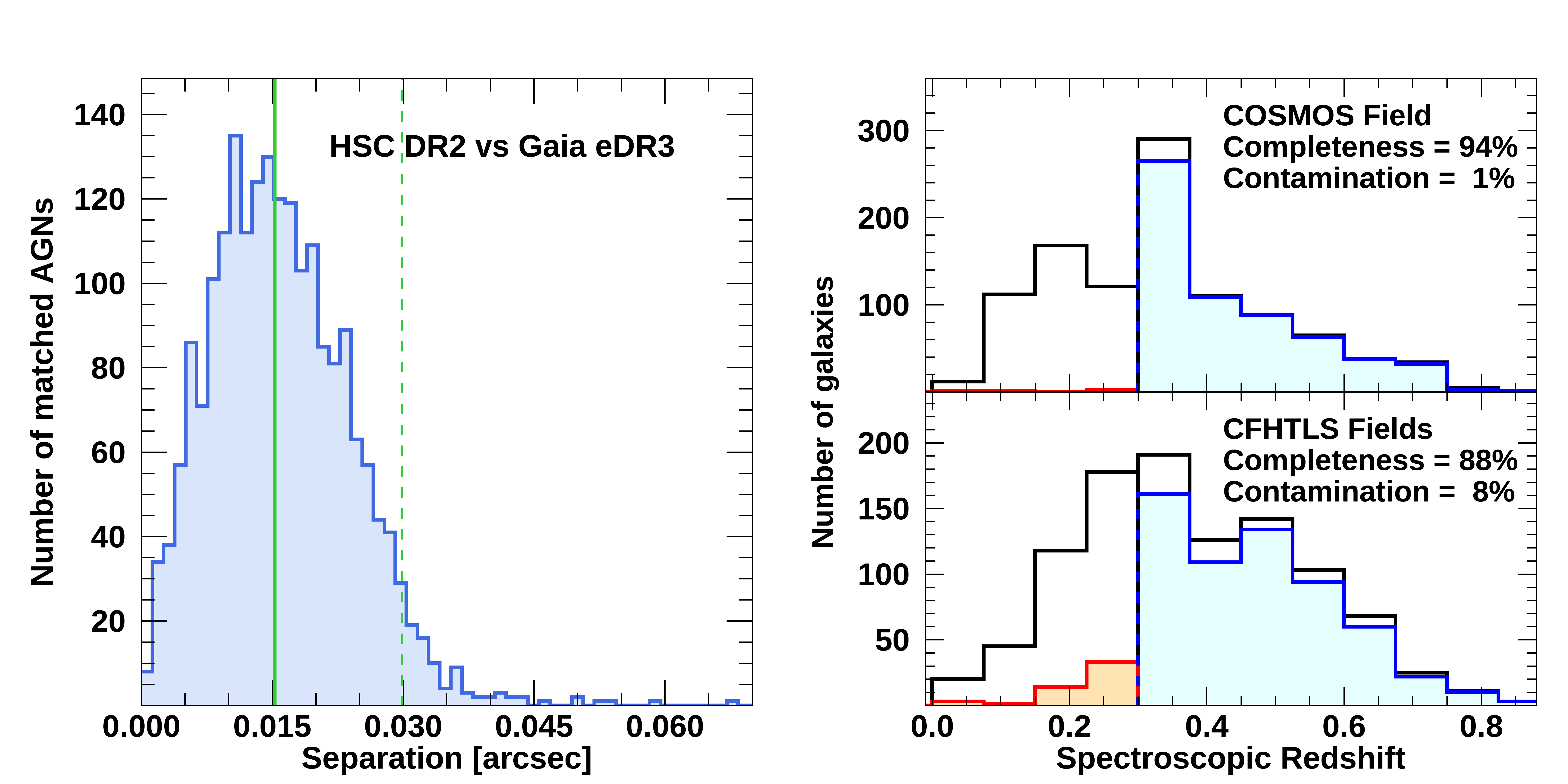} 
        \caption{Astrometric and photometric redshift quality of HSC-SSP data in the four WEAVE-StePS fields. 
                Left panel: Histogram of the coordinate cross-match statistics obtained for $\sim2000$ matched AGNs, obtained by comparing their GAIA EDR3 and HSC-SSP DR2 coordinates (see text for more details). 
                The median value is the green line at a separation of $0\farcs015$, and the dashed line is the $95\%$ percentile limit, corresponding to a separation on the sky of $\sim$ $0\farcs03$.
                Right panel: Completeness and purity of our target selection, estimated using zCOSMOS and VVDS-Wide/Deep spectroscopic data. The top panel shows the COSMOS field, and the bottom panel shows the CFHTLS-W1 and CFHTLS-W2 fields, covered by HSC deep and ultradeep and HSC wide data, respectively.  
                The histogram outlined in black shows the spectroscopic redshift of all galaxies with a magnitude $I_{AB} < 20.5$. The cyan histogram shows the distribution in spec-z of galaxies whose spectroscopic and photometric redshifts, computed using the DEmP algorithm \citep{Hsieh2014}, are above the 0.3 value redshift cut-off. The orange histogram is the spectroscopic redshift distribution of the subset of galaxies whose photometric redshift is above 0.3 when the corresponding spectroscopic redshift is below this value, as shown by the x-axis. The corresponding values for completeness and contamination of the sample selected using DEmP photometric redshifts are $94/88\%$ and $1/8\%,$ respectively (see text for more details).      
        } 
        \label{fig:Figure5}
\end{figure*}

The CFHTLS-W1 wide field is well suited for observations in the fourth trimester of the year (see its centre in Table \ref{tab:fields}). 
The area we will target within the CFHTLS-W1 wide field was chosen to overlap the spectroscopic information coming from the GAMA-02 field \citep{Baldry2018}, the VIPERS-W1 field \citep{Guzzo2014, Scodeggio2018}, and partly the VIMOS VLT Deep Survey \citep[VVDS; ][]{Lefevre2013}. 
In addition to Wide HSC-SSP photometry, the area we chose is partially covered by the data of the original Spitzer Extragalactic Representative Volume Survey, SERVS \citep{Mauduit2012}, which has now been extended \citep{Lacy2021}, and by the XMM-LSS survey and the wider XMM-XXL North Survey \citep{Pierre2011, Pierre2016, Chen2018}. Further photometric coverage includes the deep CFHT Telescope U-band data \citep[CLAUDS; see][]{Sawicki2019} and data from the VIDEO survey \citep{Jarvis2013} where available.  

In all of these four fields, we retrieved the available 3.4, 4.6, 12, and 22 ${\mu}$m data from the WISE all-sky survey \citep{Wright2010, Schlafly2019} from the literature.  

\subsection{Sample selection: Astrometry, photometric redshifts quality and field visibility \label{subsec:Astrometry_Photz}} 

A fibre-fed spectrograph such as WEAVE needs excellent astrometric quality of the input target catalogues. We checked the quality of HSC-SSP astrometry in the WEAVE-StePS fields against the reference system of the GAIA EDR3 data release \citep{Gaia2021}, which is adopted by all the WEAVE surveys.   

In the GAIA EDR3 catalogue, the Gaia celestial reference frame (CRF) is defined by a table containing $\sim1.6$ million AGNs, which are used to define the Gaia CRF, and they are themselves aligned using Very Long Baseline Array (VLBA) astrometric precision coordinates with the international celestial reference frame \citep[ICRF;][]{Lindegren2021, GAIA2022}. 

For each of the four WEAVE-StePS fields (see Section \ref{subsec:FieldsChoice}), we cross-correlated the positions of the CRF AGN table and the HSC-SSP catalogues for sources in the magnitude range $18.0 \leq I_{AB} \leq 22.0$. 
These limits avoid the brightest objects, which are probably saturated and thus have problematic centring, while stopping at reasonable magnitude limits for fainter targets.  

\begin{figure*}
        \includegraphics[width=180mm]{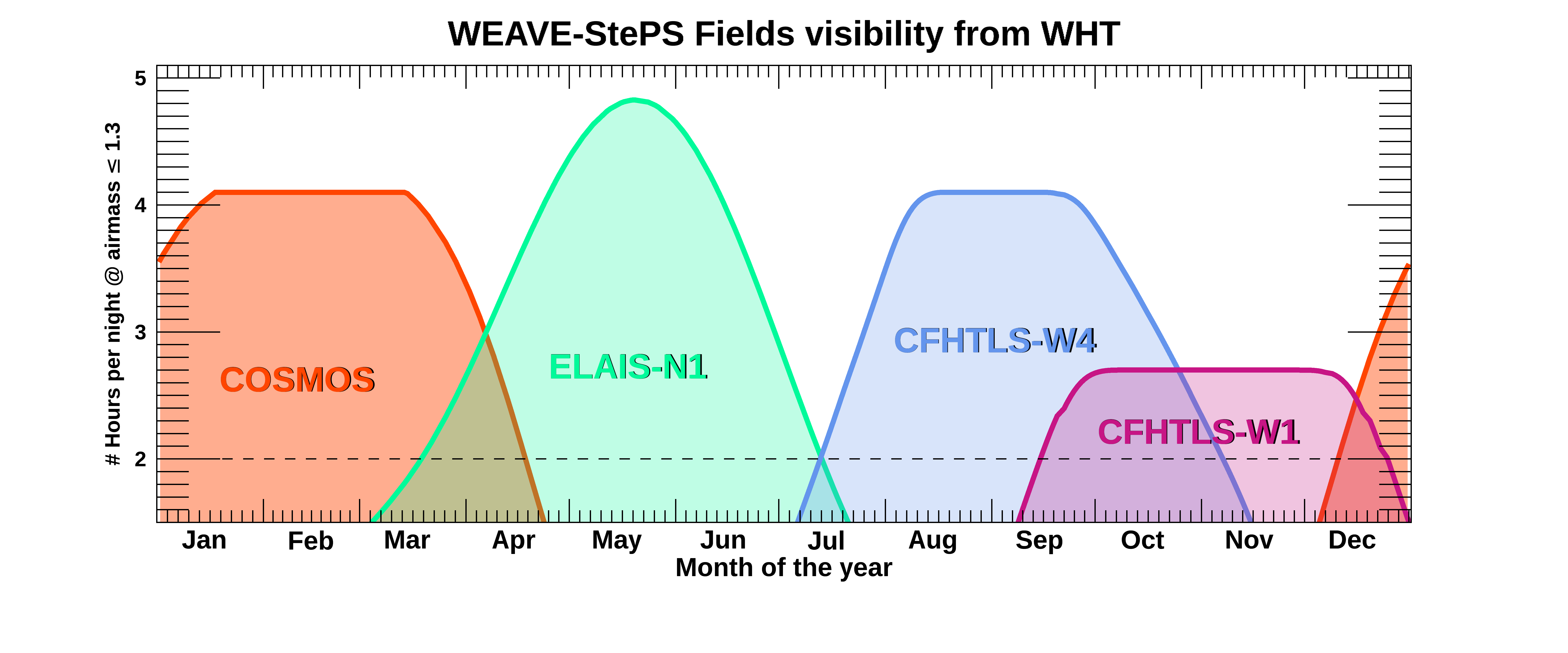} 
        \caption{Number of hours per night at an airmass below 1.3 for each 
                of the planned WEAVE-StePS fields (as labelled) throughout the year.    
        } 
        \label{fig:Figure6}
\end{figure*}

The distribution of the projected separations between the positions for the  $\sim2000$ cross-matched AGNs are presented in the left  panel of Fig.~\ref{fig:Figure5}, and the results are satisfactory for our goals: $95\%$ of the cross-matched AGNs are at separations on the sky smaller than $0\farcs03$, which is well within the expected accuracy in fibre positioning \citep[$\sim0\farcs2$ rms; see][]{Hughes2020}.

The WEAVE-StePS target selection uses a redshift cut-off, and accurate spectroscopic redshifts are available only for a subset of the sources in the WEAVE-StePS fields.  Therefore, the success of target selection relies on the accuracy of photometric redshift in our fields.  

The DR2 of HSC-SSP provides two photometric redshift values for each source that are computed with two different algorithms \citep{Tanaka2018, Nishizawa2020}.  
The first algorithm, DEmP, is based on an empirical quadratic polynomial photometric redshift fitting code \citep{Hsieh2014}, and the second algorithm, MIZUKI, is based on template fitting \citep[for more details, see][]{Tanaka2018}. 

We decided to use DEmP for our selection because it performs better ($\sigma_z/(1+z) \leq 0.03) $ at the brighter magnitudes we are interested in (down to $I_{AB} \leq 20.5$; see Table 1 and the second panel from the top of Figure 3 in \citet{Nishizawa2020}). 

As an independent test of how well DEmP performs for target selection, we used spectroscopic redshifts available in the COSMOS, CFHTLS-W4, and CFHTLS-W1 fields. In the COSMOS field, the zCOSMOS survey is available \citep{Lilly2007, Lilly2009}. This redshift survey defined its targets using only a $I_{AB} < 22.5$ magnitude limit. In the fields CFHTLS-W4 and CFHTLS-W1, the VVDS-Wide and Deep surveys are available, with targets defined using only a $I_{AB} < 22.5/24.0$ magnitude limit selection, respectively \citep[see][]{Lefevre2005, Garilli2008, Lefevre2013}. The deeper magnitude used to select spectroscopic targets for these surveys and the absence of any further colour/redshift limit enables us to extract spectroscopic redshifts for a complete subsample at $I_{AB} < 20.5$ for each of these fields to verify the purity and completeness of the photometric redshift selection based on DEmP. 

The results of this test, using a total of 2110 galaxies with spectroscopic redshifts (252, 810, and 1048 in the W1, W4, and COSMOS fields, respectively), are shown in the right panel of Fig.~\ref{fig:Figure5} for the COSMOS field and the two CFHTLS fields separately. The COSMOS field, being covered by Deep and UltraDeep HSC data, has higher-quality photometric redshifts and thus displays excellent values of completeness, roughly $\sim 94\%$, and a negligible contamination of $\sim 1\%$, when adopting DEmP for photometric redshifts selection. Similar values are expected for the ELAIS-N1 field, which is similarly covered by Deep HSC data.  The values do not change dramatically when the same analysis is performed for the two CFHTLS fields, which are mostly covered by HSC Wide photometric data. The value for completeness decreases to 88$\%$, and that of contamination rises to 8$\%$.

These values will further improve when the actual target selection is finalised, see sect.~\ref{subsec:selection}, as a non-negligible percentage of targets in our fields already possess a reliable spectroscopic redshift estimate (roughly $\sim 25\%$; see the numbers in Table~\ref{tab:fields_tiling}). 

Finally, in Fig.~\ref{fig:Figure6} we show the number of hours available at airmass lower than $1.3$ as a function of the time of year (i.e. at zenith distance smaller than $40$ degrees, the adopted constraint for our planned observations) for each WEAVE-StePS field. During all months, there are always at least 2 hours of observability for at least one of our fields. Even the field CFHTLS-W1, which is located furthest south (see Table ~\ref{tab:fields}), will be visible from WHT for $\text{about three}$ months with more than 2 hours per night at an airmass value lower than $1.3$. 


\section{Survey strategy: Field tiling and planned exposure times \label{sec:WeaveObs}}
\begin{table*}
        \caption{List of WEAVE-StePS pointing centres for each field. 
 Columns 1, 2, and 3 list the field name and RA and Dec of its centre. Column 4 lists the number of potential WEAVE-StePS targets within each pointing, N$^{\rm Targets}$ , and in parentheses, the number of targets in common (i.e. in the overlapping area) with the adjacent pointings.  Column 5 lists for each pointing the number of targets with good-quality spectroscopic redshift from the literature, N$^{\rm Targets}_{\rm spec-z}$ , and in parentheses, the number of spectroscopic redshifts common to the adjacent pointing(s). }
        \centering{
                \begin{tabular}{ccccc}
                        \hline \hline
                        Tile Name   & RA(J2000) &  Dec(J2000) &   N$^{\rm Targets}$   &  N$^{\rm Targets}_{\rm spec-z}$  \T \B  \\        
                        \hline 
                        COSMOS\_01       & 10 00 27.84   &  +02 12 21.60  &  8827 &  3892  \T \B \\
                        \hline 
                        ELAIS-N1\_01 & 16 11 00.72   &  +54 17 31.20 &  6619 (1300)  & 821 (206)   \T \\
                        ELAIS-N1\_02 & 16 11 00.72   &  +55 41 49.20 &  6210 (1300) & 589 (206)  \\
                        \multicolumn{3}{c}{Total number of distinct targets in ELAIS-N1}  & 11529   & 1204  \T \B \\
                                        \hline 
                        CFHTLS-W4\_01 & 22 10 50.16 &  +01 37 12.00 & 5351 (2363)  & 1090 (531)     \T \\
                        CFHTLS-W4\_02 & 22 14 50.40 &  +01 37 12.00 & 5822 (2363, 2137) & 1275 (531, 489)  \\
                        CFHTLS-W4\_03 & 22 18 50.40 &  +01 31 12.00 & 5989 (2137) &        1049 (489) \\
                        \multicolumn{3}{c}{Total number of distinct targets in CFHTLS-W4} & 12662  & 2394 \T \B \\
                                        \hline 
                        CFHTLS-W1\_01 & 02 17 36.00 & $-$05 15 00.00 &  5662 (2303) & 1676 (706)  \T  \\
                        CFHTLS-W1\_02 & 02 21 36.96 & $-$05 15 00.00 &  6050 (2303, 2531) & 1758 (706, 769)   \\
                        CFHTLS-W1\_03 & 02 25 37.92 & $-$05 09 00.00 & 6244 (2531, 2394) & 1886 (769, 712)    \\ 
                        CFHTLS-W1\_04 & 02 29 38.40 & $-$05 09 00.00 & 6354 (2394) & 1791 (712)  \\
                      \multicolumn{3}{c}{Total number of distinct targets in CFHTLS-W1} & 17082 & 4924 \T \B \\
                                        \hline 
      \multicolumn{3}{c}{Total number of distinct targets in all fields} & 50100 &  12414 \T \B \\
        \hline 
                \end{tabular}
        }
        \label{tab:fields_tiling}
\end{table*}

\begin{figure*}
        \includegraphics[width=180mm]{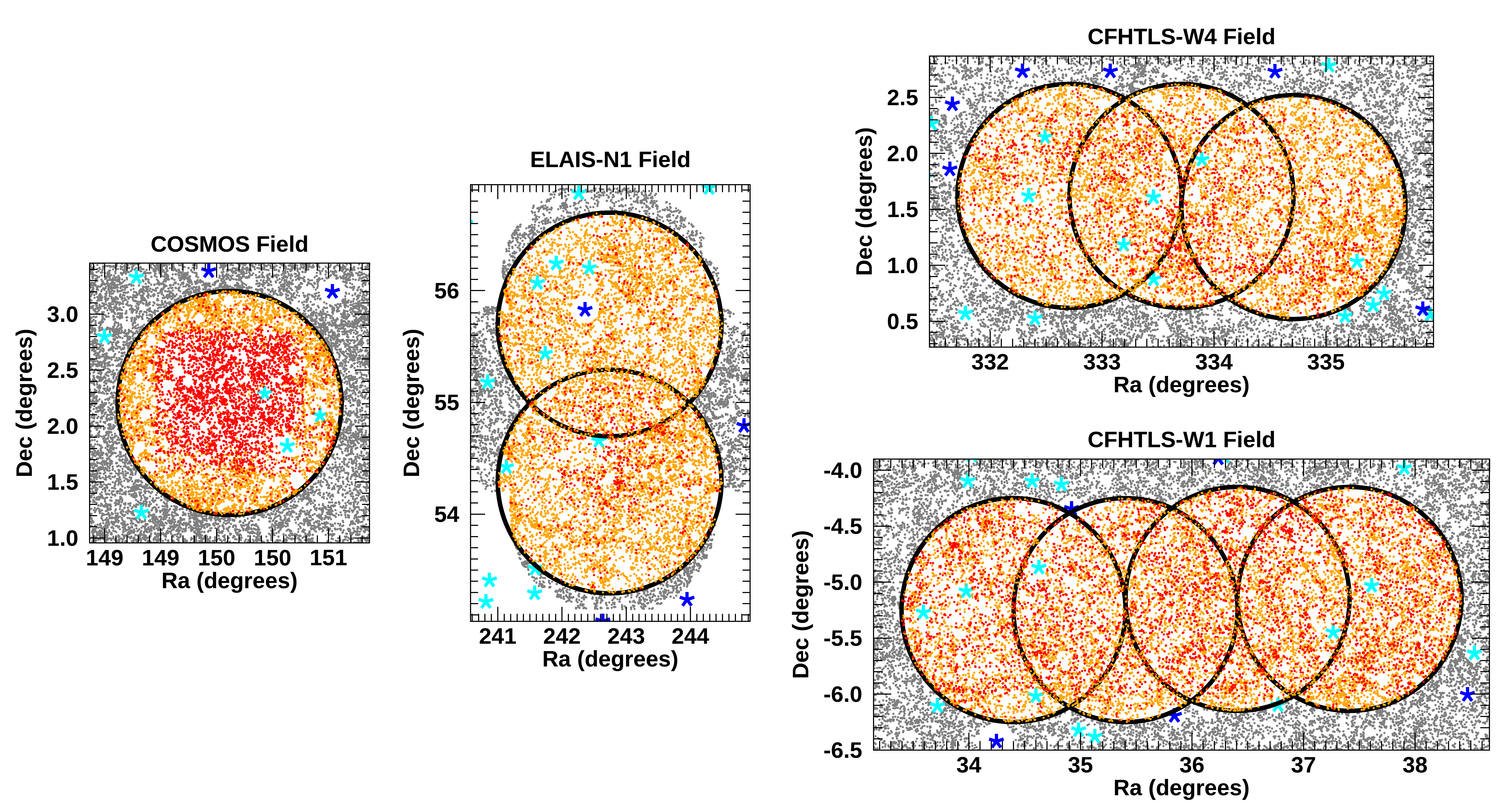} 
        \caption{Pointing layout of each field targeted by WEAVE-StePS. The orange points are the potential WEAVE-StePS targets, and the red points are the subset with available good-quality spectroscopic redshifts. The asterisks indicate the position of bright stars from GAIA catalogue: blue for those brighter than 7.5mag in the GAIA G band, and cyan for those in the GAIA G-band magnitude range 7.5-8.5. See text for more details. 
          }
        \label{fig:Figure7}
\end{figure*}

The tiling strategy adopted within each WEAVE-StePS field is driven by the goal to reach the most overlap with spectroscopic data, if available (or with good-quality HSC deep/ultra-deep photometry in the case of ELAIS-N1, where spectroscopic data are scarce). Another point we considered is the need to avoid regions in which bright stars (brighter than seventh magnitude) may hamper the quality of our data, both for the input catalogues, which will present large masked areas, and for the WEAVE spectroscopic data, which might be contaminated by stray light from internal reflections. 

The centres of each of the WEAVE-StePS pointings within each field are listed in Table \ref{tab:fields_tiling}. The table shows for each pointing the total number of targets, those with good-quality spectroscopic redshifts, and, in parentheses, those in common with adjacent pointings. The definition of good-quality spectroscopic redshifts is based on the information provided by the spectroscopic surveys available in each field. The relevant information for the main spectroscopic surveys that enter our compilation is the following. For SDSS spectroscopic redshifts, we used the three flags values ZWARNING=0, PLATEQUALITY$\neq$ ”bad”, and SPECPRIMARY=1. For GAMA redshifts, we used only spectral information flagged IS\_BEST and $nQ>2$. Finally, for VVDS, VIPERS, and zCOSMOS, we used only redshift values with spectroscopic redshift quality flag $\geq2$.   For each field (except for the COSMOS unique pointing), the final row lists the number of distinct targets within the total field area and the subset with available good-quality spectroscopic redshifts. The pointing layout in each field is shown in Fig.~\ref{fig:Figure7}, where the orange points are the potential WEAVE-StePS targets, and the red points are the subset with available good-quality spectroscopic redshifts. The asterisks indicate the position of bright stars from GAIA catalogue: blue for those brighter than 7.5mag in the GAIA G band, and cyan for those in the GAIA G-band magnitude range 7.5-8.5. Multiple pointings located within the same field overlap slightcause of the one-degree radius FoV of WEAVE.    
   
The total area covered by our survey is $\sim 25$ square degrees. The corresponding large volume covered will enable us to trace a wide range of environments at each redshift, from voids to groups and clusters, beating down cosmic variance. 

In the COSMOS field alone, as discussed in Sect.\ref{subsubsec:COSMOS}, will we extend our target selection to lower redshifts (down to $z \gtrsim 0.1$). This explains the higher number of galaxy targets within the COSMOS unique pointing in Table~\ref{tab:fields_tiling}.  For all of the other pointings, we adopted the general WEAVE-StePS selection in redshift. 

To complete the definition of our survey strategy, we quantified the exposure times needed to obtain spectra with the $S/N$ values expected to achieve our science goals ($S/N \sim 10 {\AA}^{-1}$, see Sect.~\ref{sec:GenGoals}). A simple exposure-time calculator was provided to the WEAVE team and was made available on the WEAVE web pages \footnote{see \url{https://ingconfluence.ing.iac.es/confluence/display/WEAV/Exposure+Time+Calculator}}. A more complex operation rehearsal within the WEAVE project was also performed, which used the WEAVE simulator for thorough end-to-end testing of the data-processing system \citep[see][]{Dalton2016b, Jin2022}. 
Both methods need realistic in-fibre spectral fluxes and sky-brightness values as input and provide similar outputs.  
In the following, we present the results obtained by operation rehearsal.

\begin{figure}
        \includegraphics[width=90mm]{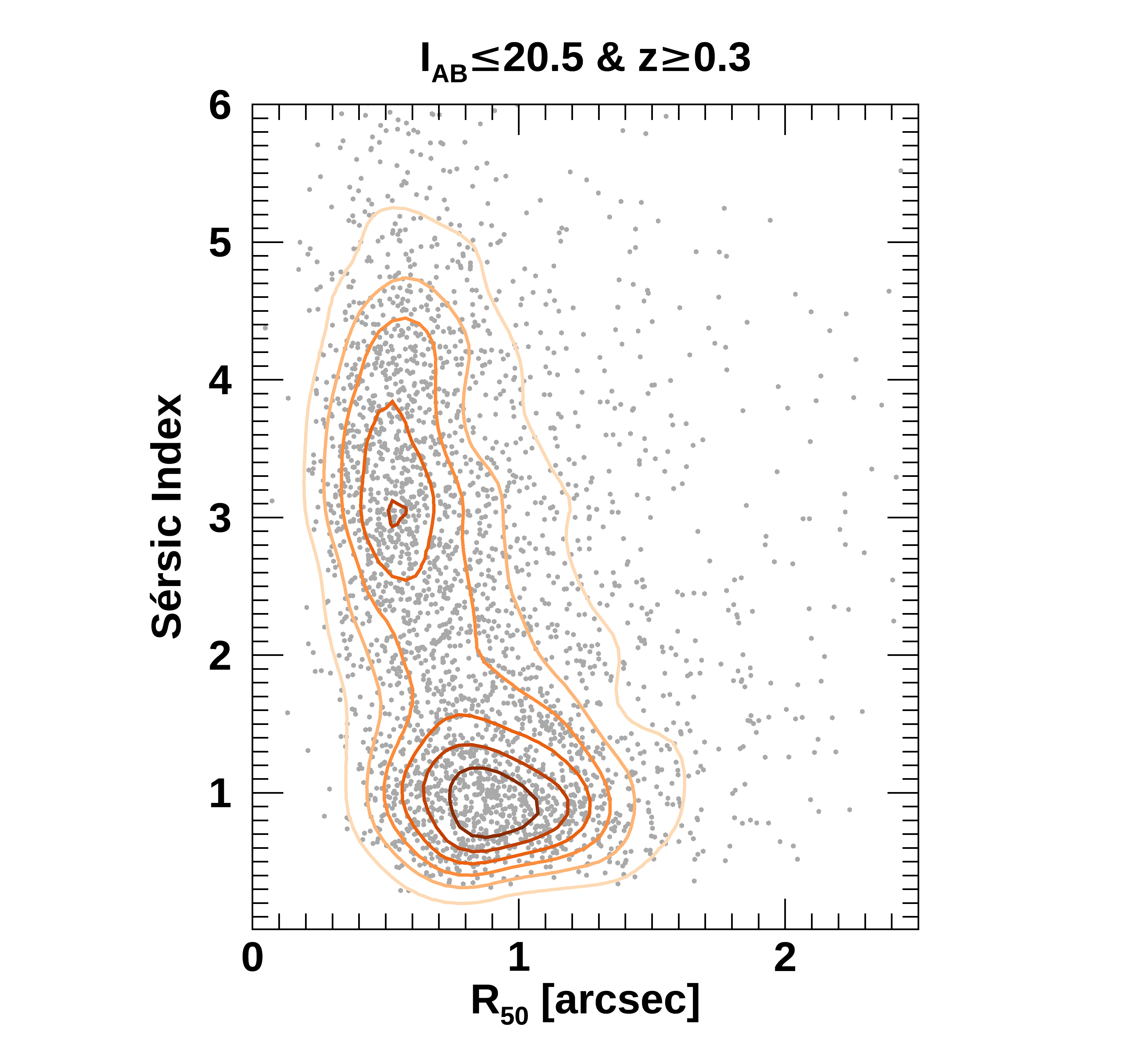} 
        \caption{Distribution of the two morphological parameters S\'ersic index and effective radius $R_{50}$  for WEAVE-StePS targets.  The points 
        were obtained from galaxies in the COSMOS field, where structural parameters for each potential target are available from HST data \citep{Sargent2007, Scarlata2007}. 
        } 
        \label{fig:Figure8}
\end{figure} 

In the COSMOS field, available HST images provide accurate information on the structural parameters \citep{Sargent2007, Scarlata2007} for each galaxy selected using WEAVE-StePS criteria, and the distribution of S\'ersic index and effective radius $R_{50}$ is shown in Fig.~\ref{fig:Figure8}.  These two parameter values were therefore used to estimate the expected in-fibre magnitude for each potential target in this field, that is, the flux loss due to the structural parameters of the source, considering only a single-component fit. This is a good first-order approximation for our tests.  By comparing the effective radius $R_{50}$ values with the 1\farcs3 WEAVE fibre diameter on the sky, it is clear that a significant percentage of the observed galaxies will suffer from non-negligible light loss due to the finite fibre size.  
We then considered a slightly pessimistic seeing of $1\farcs30$ for our simulated observations (the median seeing in La Palma varies at the zenith from $0\farcs80$ in the period April--November to $1\farcs07$ in the period December--March \footnote{\url{https://www.ing.iac.es//Astronomy/observing/conditions/#seeing}}), obtaining for each target the further flux loss due to the effect of seeing.  

We used the colour information provided by the UltraVISTA catalogue \citep{McCracken2012, Ilbert2013} to define the most appropriate E-MILES spectral template \citep{Vazdekis2016} for each target galaxy. This is to be used in the simulations so as to have the correct spectral flux distribution. Finally, we adopted a sky surface brightness in the V band of $\sim21.6$ mag arcsec$^{-2}$ for our simulations, which agrees with the typical value we expect for our fields if they are observed at an airmass limit of $\sim1.3$.  

\begin{figure*}
  \includegraphics[width=180mm]{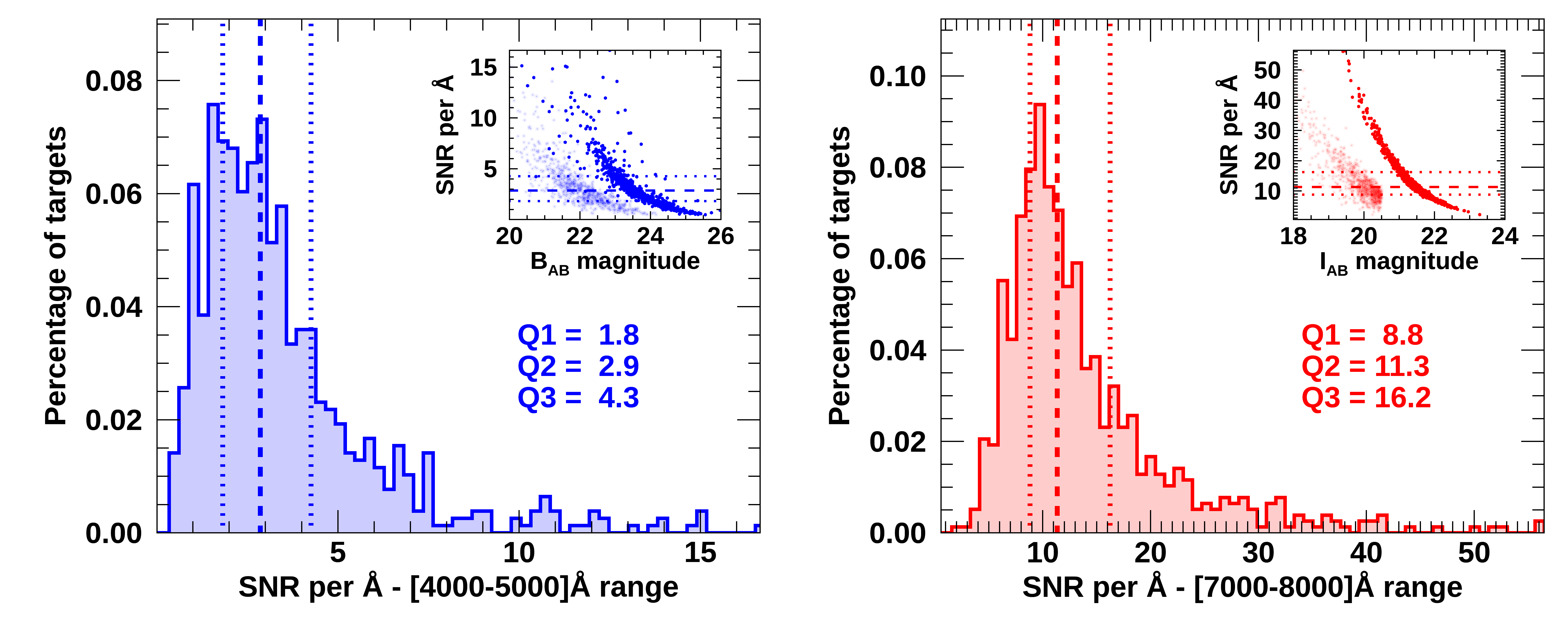}
  
  \caption{Normalized distribution of the expected $S/N$ $\AA^{-1}$ in the observed [4000-5000] and [7000-8000] ${\AA}$ wavelength ranges (left and right), assuming the stacking of 21 exposures of $20$ minutes each (including overheads) for a total exposure time of $7$ hours. The estimate was obtained using the results of the latest operation rehearsal performed within the WEAVE project. Median and quartile values are shown by the vertical lines. The inset plots in each panel show the relation between the $B_{AB}/I_{AB}$ magnitudes and observed $S/N$ $\AA^{-1}$ in the defined wavelength range. For each observed galaxy, lighter points show total magnitudes and darker points show magnitudes within the fibre, see text for more details. 
 See text for more details.             
        } 
        \label{fig:Figure9}
\end{figure*}

Based on a realistic distribution of $S/N$ $\AA^{-1}$ for a representative subset of our targets, roughly $\sim 800$ in number, we then determined the optimal choice in exposure time to obtain the I-band target goal of $S/N \sim 10$ $\AA^{-1}$.    

The left and right panels of Fig.~\ref{fig:Figure9} show the normalized distribution of the expected $S/N$ $\AA^{-1}$ in the observed [4000-5000] and [7000-8000] ${\AA}$ wavelength ranges, respectively, assuming the stacking of 21 exposures of $20$ minutes each (including overheads), for a total exposure time of $7$ hours. 

Planning an exposure time of $\sim7$ hours for each WEAVE-StePS pointing should thus enable us to reach the required $S/N \sim 10 {\AA}^{-1}$ in the observed I band for the majority of the observed galaxies, even in the slightly pessimistic observing conditions we considered.  The median $S/N$ $\AA^{-1}$ in the observed I band is $\sim11.3$, with lower/upper quartile values of $\sim8.8$/$\sim16.2$. This means that fewer than $\sim 25\%$ of the observed spectra will be at an $S/N$ that is significantly below our fiducial value of $10 {\AA}^{-1}$ in the observed I band. These spectra will be analysed using stacking techniques within homogeneous bins of galaxy properties (stellar mass, redshift, velocity dispersion, etc.).  The corresponding $S/N$ values in the blue arm are significantly lower, as expected. However, as discussed in \citet{Costantin2019} and Ditrani et al. (2023, {\it in prep.}), these values are sufficient for our analysis.

According to the allocation of fibre-hours foreseen for WEAVE-StePS within the general WEAVE survey, and assuming 7 hours of exposure time for each pointing, we can expect to observe a total of 35 WEAVE-StePS pointings.

To compute the total number of galaxies observed by WEAVE-StePS at the end of the five-year survey campaign, we need a rough estimate of the number of galaxies that is observed at each pointing. 

Out of the $\sim950$ fibres available per pointing in MOS mode, $\sim100$ sky-fibres will be allocated to observe empty-sky regions. These fibres will be used to perform an accurate sky subtraction. A further $\sim10$ fibres are devoted to observations of white dwarf stars as part of the WEAVE calibration plan and of a dedicated WEAVE survey. 

For an a posteriori sanity check of the quality of the sky subtraction, we plan to allocate $\sim30$ fibres to empty sky regions in each pointing, the {\it undeclared sky-fibres}. A further $\sim20$ fibres will be reserved for known star calibrators that are available within each field to verify the flux calibration. These additional calibrators were selected from the GAIA EDR3 database and were matched with the SDSS-DR16 spectroscopic and photometric catalogues. The majority are type A or F stars with Gaia BP/RP magnitudes in the range $14-19$, and high-quality spectra are available from SDSS ($S/N \gtrsim 30$ in $r-$band).  

We finally need to consider the overall fibre allocation inefficiencies (e.g. parked and broken fibres) and the possibility of some residual contamination of our input sample by stars and/or lower-redshift galaxies.  A realistic estimate of the mean WEAVE-StePS targets observed in each pointing is thus $\sim700$, which agrees well with the results of tests using the code Configure \citep{Terrett2014, Hughes2022}. This software will be used to position fibres during actual survey operations. As a result, we expect that a total of $\sim25,000$ galaxies are observed by WEAVE-StePS by the end of the five-year survey campaign. 

Given the distribution of expected $S/N$ $\AA^{-1}$ for our targets as shown in Fig.~\ref{fig:Figure8}, such a large sample implies that WEAVE-StePS will produce a significant number of galaxy spectra ($\sim 3500$) at $S/N \geq 20$, which is comparable to the values quoted for the LEGA-C survey \citep{Straatman2018}, whose spectra are at a slightly lower resolution of $R \sim 2500$ in a narrower observed spectral wavelength range [6000 - 8800] $\AA$. Last but not least, and as already mentioned, the large total area covered by our survey will enable us to observe a wide range of environments at each redshift, reducing the impact of cosmic large-scale structure variance. 

We note from Table~\ref{tab:fields_tiling} that the number of distinct passes we may perform in each pointing, given the total number of targets available per pointing,  varies from a maximum of $\sim$7-8, as in the case of COSMOS, down to a minimum of $\sim3$ as in the case of  CFHTLS-W1/CFHTLS-W4, where each chosen area overlaps non-negligibly with adjacent areas. ELAIS-N1 is an intermediate case for which we may easily consider positioning $\text{about four}$ passes per each pointing. The WEAVE-StePS planning will therefore easily fit in the general WEAVE survey planning, with a roughly homogeneous distribution of our targeted areas during the different months of the year. 

\section{Summary and conclusions \label{sec:Conclusions}}

The WEAVE spectrograph mounted on the WHT has unprecedented multiplex capabilities in MOS\ mode over a wide field of view (950 fibres over $\sim$3 square degrees). These characteristics, in combination with its wide spectral coverage (3700 to 9600 \AA) in the low-resolution mode ($R \sim$ 5000, a factor of $\sim$ 10 improvement of typical literature values), will enable astronomers to carry out surveys that have not been feasible so far. 
In this paper, we presented the WEAVE-StePS survey, one of the five extragalactic surveys that will be carried out with WEAVE and whose observations are expected to start in the first trimester of 2023. We have outlined the main scientific objectives of StePS, the detailed survey design and strategy, and the expected spectral and sample characteristics. 

WEAVE-StePS will collect $\sim$ 25000 spectra of galaxies. The vast majority of them is located in the intermediate-redshift range between LEGA-C and SDSS: 0.3$ \leq z \leq$ 0.7. This will characterise in detail the properties and stellar histories of a large sample of galaxies during this cosmic period for the first time. Only $\sim 4\%$ of the targeted sample will be at $z \ge$0.7, and the lower redshift cut-off will be lowered to $z \geq$ 0.1 in the COSMOS field alone. 
The planned long exposure times (7 hours), combined with the WEAVE throughput, will enable WEAVE-StePS to gather high-quality spectra ($S/N \geq 10\,\AA^{-1}$ in the I band for the majority of the spectra, with a resolution $R \sim$ 5000) and to accurately measure for individual galaxies the star formation rates, histories, and timescales, gas and stellar metallicities, stellar and dynamical masses, gas inflows and outflows, and AGN/LINER emission.   
The $\sim 25$ square degrees of the area covered by WEAVE-StePS in four of the most well-studied extragalactic fields (COSMOS, ELAIS-N1, CFHTLS-W4, and CFHTLS-W1, all with extensive ancillary data) will include all types of galaxy environments, from voids to filaments, galaxy groups, and clusters, thus providing the opportunity to study galaxy evolution at these epochs for different environmental conditions and in a wide range of dark matter halo masses. 

WEAVE-StePS will go back in cosmic time beyond SDSS and will fill the crucial redshift interval that bridges the gap between LEGA-C and SDSS data, enabling us to continuously retrace the evolutionary path of galaxies over the $\sim 8$ Gyr since $z \sim 1$.

%
\begin{acknowledgements}  

  Funding for the WEAVE facility has been provided by UKRI STFC, the University of Oxford, NOVA, NWO, Instituto de Astrofísica de Canarias (IAC), the Isaac Newton Group partners (STFC, NWO, and Spain, led by the IAC), INAF, CNRS-INSU, the Observatoire de Paris, Région Île-de-France, CONCYT through INAOE, Konkoly Observatory (CSFK), Max-Planck-Institut für Astronomie (MPIA Heidelberg), Lund University, the Leibniz Institute for Astrophysics Potsdam (AIP), the Swedish Research Council, the European Commission, and the University of Pennsylvania.

  The WEAVE Survey Consortium consists of the ING, its three partners, represented by UKRI STFC, NWO, and the IAC, NOVA, INAF, GEPI, INAOE, and individual WEAVE Participants. Please see the relevant footnotes for the WEAVE website\footnote{\url{https://ingconfluence.ing.iac.es/confluence//display/WEAV/The+WEAVE+Project}} and for the full list of granting agencies and grants supporting WEAVE\footnote{\url{https://ingconfluence.ing.iac.es/confluence/display/WEAV/WEAVE+Acknowledgements}}.

  This work makes use of data from the European Space Agency (ESA) mission {\it Gaia} (\url{https://www.cosmos.esa.int/gaia}), processed by the {\it Gaia}
Data Processing and Analysis Consortium (DPAC, \url{https://www.cosmos.esa.int/web/gaia/dpac/consortium}). Funding for the DPAC has been provided by national institutions, in particular, the institutions participating in the {\it Gaia} Multilateral Agreement.

This work makes use of data from the VIMOS VLT Deep Survey, the VIPERS-MLS database and the HST-COSMOS database, operated by CeSAM/Laboratoire d'Astrophysique de Marseille, France.

This work makes use of observations obtained with MegaPrime/MegaCam, a joint project of CFHT and CEA/IRFU, at the Canada-France-Hawaii Telescope (CFHT) which is operated by the National Research Council (NRC) of Canada, the Institut National des Science de l'Univers of the Centre National de la Recherche Scientifique (CNRS) of France, and the University of Hawaii.

This work makes use of data products produced at Terapix available at the Canadian Astronomy Data Centre as part of the Canada-France-Hawaii Telescope Legacy Survey, a collaborative project of NRC and CNRS.

 R.G.B. and R.G.D. acknowledge financial support from the grants CEX2021-001131-S funded by MCIN/AEI/10.13039/501100011033, SEV-2017-0709, and to PID2019-109067-GB100. 

A.F.M. acknowledges financial support from grant CEX2019-000920-S from the Spanish Ministry of Science and Innovation.

M.Bi. acknowledges support from STFC grant numbers ST/N021702/1

C.S. is supported by a 'Hintze Fellow' at the Oxford Centre for Astrophysical Surveys, which is funded through generous support from the Hintze Family Charitable Foundation.  

G.B., M.Bo., F.R.D., A.I., F.L.B., M.L., P.M., B.P. C.T. D.V. and S.Z. acknowledge financial support from INAF funds, program 1.05.01.86.16 - Mainstream 2019.   

L.C. acknowledges financial support from Comunidad de Madrid under Atracci\'on de Talento grant 2018-T2/TIC-11612. 

J.A. acknowledges financial support from INAF-WEAVE funds, program 1.05.03.04.05 and INAF-OABrera funds, program 1.05.01.01. 

J.H.K. acknowledges financial support from the State Research Agency (AEI-MCINN) of the Spanish Ministry of Science and Innovation under the grant "The structure and evolution of galaxies and their central regions" with reference PID2019-105602GB-I00/10.13039/501100011033, and from the ACIISI, Consejer\'{i}a de Econom\'{i}a, Conocimiento y Empleo del Gobierno de Canarias and the European Regional Development Fund (ERDF) under grant with reference PROID2021010044.

A special thanks to Daniela Bettoni for her suggestions and comments and to the anonymous referee for his/her useful comments.

\end{acknowledgements}


\bibliographystyle{aa}
\bibliography{WEAVE_StePS_AandA}


\end{document}